\pdfoutput=1
\documentclass{JHEP3}
\usepackage{amsmath}
\usepackage{graphics}
\usepackage{epsfig}

\def\({\left(}
\def\){\right)}
\def\[{\left[}
\def\]{\right]}
\def\<{\langle}
\def\>{\rangle}
\def\ap{\alpha}
\def\bt{\beta}
\def\gm{\gamma}
\def\de{\delta}
\def\sg{\sigma}

\newcommand{\be}{\begin{equation}}
\newcommand{\ee}{\end{equation}}
\newcommand{\bal}{\begin{aligned}}
\newcommand{\eal}{\end{aligned}}

\newcommand{\labell}[1]{\label{#1}}

\title{4-particle Amplituhedron at 3-loop and its Mondrian Diagrammatic Implication}

\author{Junjie Rao$^{ab}$\footnote{Email: jrao@aei.mpg.de}\\
{$^a$Max Planck Institute for Gravitational Physics (Albert Einstein Institute), 14476 Potsdam, Germany}\\
{$^b$Zhejiang Institute of Modern Physics, Zhejiang University, Hangzhou, 310027, P. R. China}}

\abstract{This article provides a direct calculation of the 4-particle amplituhedron at 3-loop order, by introducing
a set of practical tricks. After delicately rearranging each piece of this calculation, we find a
suggestive connection between positivity conditions and Mondrian diagrams, which will be quantitatively defined.
Such a pattern can be generalized for all Mondrian diagrams among all those contribute to the 4-particle integrand
of planar $\mathcal{N}\!=\!4$ SYM to all loop orders, as a subsequent work 1712.09994 will show.}


\begin{document}
\maketitle

\section{Introduction}

The amplituhedron proposal for 4-particle integrand of planar $\mathcal{N}\!=\!4$ SYM to all loop orders
\cite{Arkani-Hamed:2013jha,Arkani-Hamed:2013kca,Franco:2014csa,Galloni:2016iuj} is a novel reformulation
which only uses positivity conditions for all physical poles to construct the loop integrand.
At 2-loop order, as the first nontrivial case, we have just one (mutual) positivity condition
\be
D_{12}\equiv(x_2-x_1)(z_1-z_2)+(y_2-y_1)(w_1-w_2)>0, \labell{eq-1}
\ee
where $x_i\!=\!\<A_iB_i\,14\>$, $y_i\!=\!\<A_iB_i\,34\>$, $z_i\!=\!\<A_iB_i\,23\>$, $w_i\!=\!\<A_iB_i\,12\>$
and $D_{ij}\!=\!\<A_iB_i\,A_jB_j\>$ are all possible physical poles in terms of momentum twistor contractions,
and $x_i,y_i,z_i,w_i$ are trivially set to be positive for the $i$-th loop. The resulting integrand is the double-box
topology of two possible orientations, and it is symmetrized for two sets of loop variables \cite{Arkani-Hamed:2013kca}.
As the loop order increases, its calculational complexity grows explosively due to the highly nontrivial intertwining of
all $L(L\!-\!1)/2$ positivity conditions of $D_{ij}$'s. As far as the 3-loop case, it is done under significant
simplification brought by double cuts \cite{Arkani-Hamed:2013kca}, still there is considerable complexity that obscures
its somehow simple mathematical structure, as we will reveal in this article and the subsequent work \cite{An:2017tbf}.

As an illuminating appetizer, we reformulate the 2-loop case in the following. As usual, let's preserve $z_1,z_2$
for imposing $D_{12}\!>\!0$, and triangulate the space spanned by $x_1,x_2,y_1,y_2,w_1,w_2$. We introduce the
\textit{ordered subspaces} characterized by, for instance:
\be
X(12)Y(12)W(12)\equiv\frac{1}{x_1(x_2-x_1)}\frac{1}{y_1(y_2-y_1)}\frac{1}{w_1(w_2-w_1)},
\ee
which is a $d\log$ form (omitting the measure factor) of the orderings $x_1\!<\!x_2$, $y_1\!<\!y_2$
and $w_1\!<\!w_2$. In this particular subspace, positivity condition $\eqref{eq-1}$ unambiguously demands
\be
z_1-z_2>\frac{y_{21}w_{21}}{x_{21}},
\ee
where $x_{21}\!\equiv\!x_2-x_1$ and so forth. Here, $x_{21},y_{21},w_{21}$ can be treated as genuinely positive variables
which replace the original $x_2,y_2,w_2$. Then the relevant $d\log$ form for $z_1,z_2$ is simply
\be
\frac{1}{z_2(z_1-z_2-y_{21}w_{21}/x_{21})}=\frac{x_{21}z_1}{z_1z_2\,D_{12}},
\ee
analogously, for $X(12)Y(12)W(21)$ we have
\be
\frac{1}{z_1}\(\frac{1}{z_2}-\frac{1}{z_2-z_1-y_{21}w_{12}/x_{21}}\)=\frac{x_{21}z_1+y_{21}w_{12}}{z_1z_2\,D_{12}}.
\ee
A seemingly farfetched observation is, after we flip $W(12)$ to $W(21)$, the additional term $y_{21}w_{12}$ appears
in the numerator above due to the orderings of $y_1,y_2$ and $w_1,w_2$ are now opposite, allowing one to orient the
double box ``vertically'', as explained diagrammatically below.

\begin{figure}
\begin{center}
\includegraphics[width=0.2\textwidth]{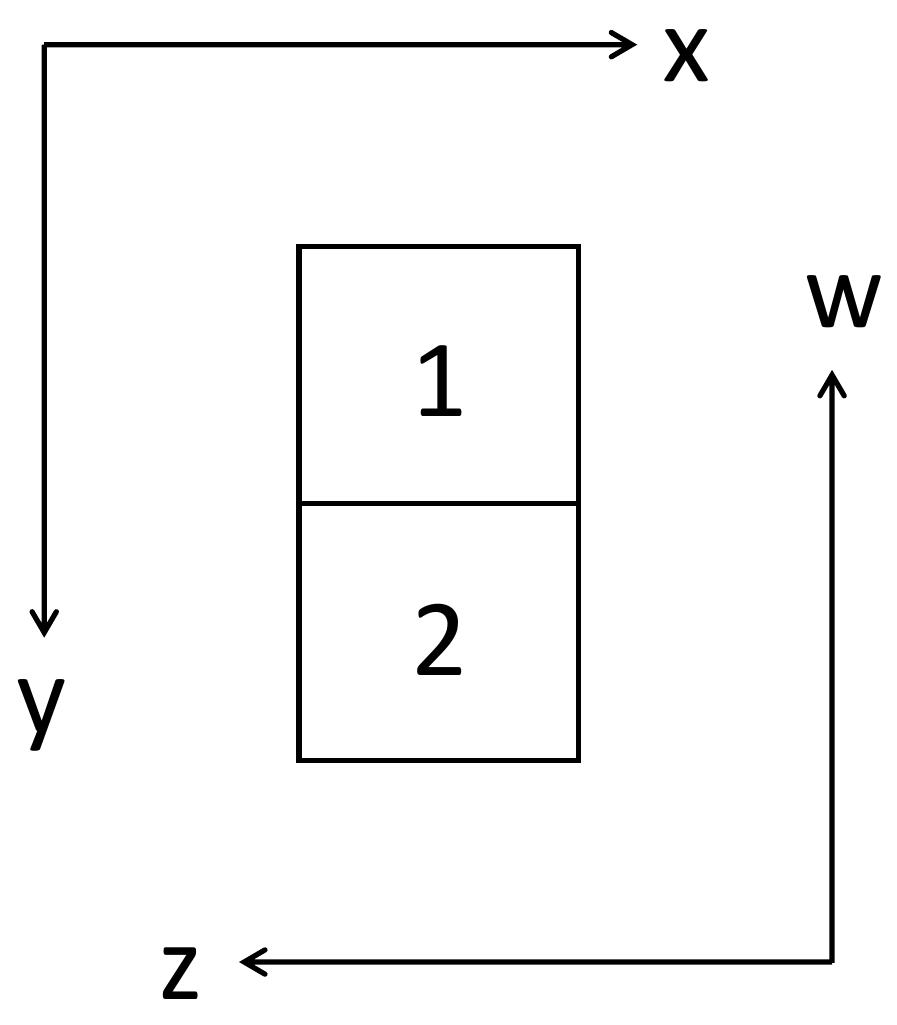}
\caption{Giving the orderings of positive variables $x_i,y_i,z_i,w_i$ a directional meaning.} \label{fig-1}
\end{center}
\end{figure}

In figure \ref{fig-1}, we have chosen two perpendicular directions for $x$ and $y$, while the $z$ and $w$ directions
are opposite to those of $x$ and $y$ respectively. Then we assign each loop with a number as usual, but now these numbers
have a meaning of orderings of positive variables. Since loop number 2 is below 1,
we naturally interpret this as $y_2\!>\!y_1$, and similarly $w_1\!>\!w_2$.
In this way, it is straightforward to conclude that, if we flip $w_1\!>\!w_2$ back to $w_2\!>\!w_1$,
there is no consistent way to place loop numbers $1,2$ vertically so the double box can be only oriented horizontally!

After we sum the numerators above over $W(12)$ and $W(21)$ respectively, namely
\be
\frac{1}{w_1(w_2-w_1)}(x_{21}z_1)+\frac{1}{w_2(w_1-w_2)}(x_{21}z_1+y_{21}w_{12})=\frac{x_{21}z_1+y_{21}w_1}{w_1w_2},
\ee
the ordering of $w_1,w_2$ is nicely wiped off as expected. Similarly, after summing all such $d\log$ forms over $y$-
and $x$-space, we get the symmetrized numerator $(x_1z_2\!+\!x_2z_1\!+\!y_1w_2\!+\!y_2w_1)$
\cite{Arkani-Hamed:2013kca}, and the corresponding denominator is $\,x_1x_2y_1y_2z_1z_2w_1w_2\,D_{12}\,$
which is free of any ordering of $x_i,y_i,z_i,w_i$.

Actually, positivity condition \eqref{eq-1} already has a diagrammatic implication. If both $(x_2\!-\!x_1)(z_1\!-\!z_2)$ and
$(y_2\!-\!y_1)(w_1\!-\!w_2)$ are positive, which means both the orderings of $x_1,x_2$ and $z_1,z_2$, and those of
$y_1,y_2$ and $w_1,w_2$, are opposite, $D_{12}$ is trivially positive. In this case both horizontal and vertical
orientations of the double box are allowed and we call them \textit{seed diagrams} with respect to a particular
ordered subspace, while those final diagrams are obtained after we sum seed diagrams over all ordered subspaces
that admit them for each distinct topology with an orientation.
If $(y_2\!-\!y_1)(w_1\!-\!w_2)$ is negative, for instance, after we switch to
ordered subspace $Y(12)W(12)$ or $Y(21)W(21)$, $D_{12}$ is then conditionally positive,
since $x_{21}z_{12}$ must be greater than $y_{21}w_{21}$ or $y_{12}w_{12}$, and the double box can be only oriented
horizontally. But if both $(x_2\!-\!x_1)(z_1\!-\!z_2)$ and $(y_2\!-\!y_1)(w_1\!-\!w_2)$ are negative, $D_{12}$
cannot be positive and diagrammatically it means there is no legal seed diagram in this ordered subspace.

One may already notice that this diagrammatic setting only makes sense when all internal lines can be oriented
either horizontally or vertically (which are also borders of adjacent loops), and consequently,
only 3- and 4-vertex are admitted. This category is called the \textit{Mondrian diagrams} \cite{{Bern:2005iz}},
as we will exactly define them based on the amplituhedron setting in the subsequent work. Inside a Mondrian diagram,
any two loops may only have a horizontal contact, vertical contact or no contact. To characterize
a Mondrian diagram in this way turns out to be sufficient for unambiguously identifying its topological structure!

However, it is known beyond 3-loop, non-Mondrian diagrams also contribute to the planar 4-particle integrand
\cite{Bern:2006ew,Bern:2007ct}. Therefore, for now we have to limit ourselves to the Mondrian types until further
generalization is available. Nevertheless, at 3-loop such a setting can perfectly connect the amplituhedron and the
actual loop integrand. To understand its motivation, we will first present the direct calculation in the context
of amplituhedron (or positivity conditions more precisely).

This article is organized as follows. Section \ref{sec2} presents the fundamentals of positive $d\log$ forms
which are necessary for posterior formulation. Section \ref{sec3} introduces the trick of intermediate variables
to handle the 4-particle amplituhedron at 3-loop as a bridge towards the precise description of imposing 3 positivity
conditions simultaneously. Section \ref{sec4} continues to sum the former results over all ordered subspaces,
as we find it ``almost'' reaches the correct answer. Section \ref{sec5} further refines co-positive products
for each ordered subspace of $y$ and $w$, based on the discussions in terms of intermediate variables,
as a precise description. Section \ref{sec6} sums the refined co-positive products by delicately separating the contributing
and the spurious parts, where the former manifest the Mondrian diagrammatic interpretation, while the latter
sum to zero at the end as their name implies.

\newpage
\section{Fundamentals of Positive $d\log$ Forms}
\label{sec2}

First, we will extend the fundamentals of positive $d\log$ forms in \cite{Arkani-Hamed:2013kca},
as the minimal techniques necessary for the posterior sections. It is known that, for a generic positive variable
ranging from $a$ to $b$ ($a\!<\!b$), its $d\log$ form is given by
\be
\frac{b-a}{(x-a)(b-x)},
\ee
in particular, for $a\!=\!0$ this becomes
\be
\frac{b}{x(b-x)},
\ee
while for $b\!=\!\infty$ it becomes
\be
\frac{1}{x-a},
\ee
finally, if $b\!=\!a$ for the two special cases above, we have
\be
\frac{1}{x-a}+\frac{a}{x(a-x)}=\frac{1}{x}, \labell{eq-2}
\ee
which will be named as the \textit{completeness relation}. It has a natural interpretation as the sum of
projective lengths of two complementary positive intervals. Note that we have treated $a$ as a constant above, while it
could also be a positive variable. In that case we only need an additional form $1/a$, so the completeness relation
now becomes
\be
\frac{1}{a(x-a)}+\frac{1}{x(a-x)}=\frac{1}{a\,x},
\ee
where the LHS characterizes nothing but two ordered subspaces in which $x\!>\!a$ and $x\!<\!a$ respectively.
A trivial generalization of \eqref{eq-2} for $n$ $x_i$'s satisfying $x_1\ldots x_n\!\gtrless\!a$ is then
\be
\frac{1}{x_1\ldots x_n-a}+\frac{a}{x_1\ldots x_n(a-x_1\ldots x_n)}=\frac{1}{x_1\ldots x_n},
\ee
here, for example, $x_1\ldots x_n\!>\!a$ is characterized by
\be
\frac{1}{x_2\ldots x_n}\frac{1}{x_1-a/(x_2\ldots x_n)}=\frac{1}{x_1\ldots x_n-a}.
\ee
Another less straightforward generalization of \eqref{eq-2} for $x_1\!+\!\ldots\!+\!x_n\!\gtrless\!a$ is
\be
\frac{x_1+\ldots+x_n}{x_1\ldots x_n(x_1+\ldots+x_n-a)}+\frac{a}{x_1\ldots x_n(a-x_1-\ldots-x_n)}
=\frac{1}{x_1\ldots x_n}, \labell{eq-3}
\ee
where both parts of the LHS can be proved recursively. If we assume they hold for $x_1\!+\!\ldots\!+\!x_{n-1}\!>\!a$
and $x_1\!+\!\ldots\!+\!x_{n-1}\!<\!a$ respectively, to obtain the form of $x_1\!+\!\ldots\!+\!x_n\!>\!a$
we must separate it into two parts as
\be
\bal
&\frac{x_1+\ldots+x_{n-1}}{x_1\ldots x_{n-1}(x_1+\ldots+x_{n-1}-a)}\frac{1}{x_n}
+\frac{a}{x_1\ldots x_{n-1}(a-x_1-\ldots-x_{n-1})}\frac{1}{x_1+\ldots+x_n-a}\\
=\,&\frac{x_1+\ldots+x_n}{x_1\ldots x_n(x_1+\ldots+x_n-a)},
\eal
\ee
where in the first line, $x_n$ is positive in the first term, but greater
than $(a\!-\!x_1\!-\!\ldots\!-\!x_{n-1})$ in the second. The sum in the second line nicely returns to the form
for $n$ $x_i$'s. To obtain the form of $x_1\!+\!\ldots\!+\!x_n\!<\!a$, we can simply insert
\be
\frac{(a-x_1-\ldots-x_{n-1})}{x_n((a-x_1-\ldots-x_{n-1})-x_n)}
\ee
into the form for $(n\!-\!1)$ $x_i$'s, note that $(a\!-\!x_1\!-\!\ldots\!-\!x_{n-1})$ is treated as one positive
variable above. These two forms, as well as completeness relation \eqref{eq-3},
are often used in the subsequent derivation.

It is also convenient to introduce the \textit{co-positive} product of forms. For example, for
$y\!>\!x_1,\ldots,x_n$, to obtain its form we can divide it into $n!$ parts with respect to $n!$ ordered subspaces
in which $x_{\sg_1}\!<\!\ldots\!<\!x_{\sg_n}$ and $\{\sg_1,\ldots,\sg_n\}$ is a permutation of $\{1,\ldots,n\}$.
Then we need to simplify
\be
I_n(X_n,y)\equiv\sum_{\sg_n}\frac{1}{x_{\sg_1}}\frac{1}{x_{\sg_2}-x_{\sg_1}}\cdots\frac{1}{x_{\sg_n}-x_{\sg_{n-1}}}
\frac{1}{y-x_{\sg_n}}
\ee
by induction, here $X_n\!=\!\{x_1,\ldots,x_n\}$. Now let's focus on $x_n$'s location in each permutation while omitting
those of $x_1,\ldots,x_{n-1}$, it is straightforward to regroup the sum in order to reach
\be
\bal
\frac{I_n}{I_{n-1}}=\,&\frac{y-x_{\sg_{n-1}}}{(x_n-x_{\sg_{n-1}})(y-x_n)}
+\frac{x_{\sg_{n-1}}-x_{\sg_{n-2}}}{(x_n-x_{\sg_{n-2}})(x_{\sg_{n-1}}-x_n)}
+\ldots+\frac{x_{\sg_2}-x_{\sg_1}}{(x_n-x_{\sg_1})(x_{\sg_2}-x_n)}+\frac{x_{\sg_1}}{x_n(x_{\sg_1}-x_n)}\\
=\,&\frac{y-x_{\sg_{n-1}}}{(x_n-x_{\sg_{n-1}})(y-x_n)}+\frac{x_{\sg_{n-1}}}{x_n(x_{\sg_{n-1}}-x_n)}=\frac{y}{x_n(y-x_n)},
\eal
\ee
therefore
\be
I_n=\frac{y^{n-1}}{x_1\ldots x_n(y-x_1)\ldots(y-x_n)}
\equiv\frac{1}{x_1(y-x_1)}\cap\cdots\cap\frac{1}{x_n(y-x_n)}, \labell{eq-4}
\ee
here the symbol $\cap$ is the co-positive product operation. This product denotes the intersected subspace
of a number of different subspaces as one form. If we evaluate the residue of $I_n$ at $y\!=\!\infty$, it returns to
\be
I_n(X_n,\infty)=\int_\infty\frac{dy}{y}\times\frac{y^n}{x_1\ldots x_n(y-x_1)\ldots(y-x_n)}=\frac{1}{x_1\ldots x_n},
\ee
which is the completeness relation of $n$ positive variables as all $x_i$'s are trivially less than infinity.

Analogously, for $y\!<\!x_1,\ldots,x_n$ we need to simplify
\be
J_n(y,X_n)
=\sum_{\sg_n}\frac{1}{\,y\,}\frac{1}{x_{\sg_1}-y}\frac{1}{x_{\sg_2}-x_{\sg_1}}\cdots\frac{1}{x_{\sg_n}-x_{\sg_{n-1}}}
\ee
with the aid of
\be
\bal
\frac{J_n}{J_{n-1}}
=\,&\frac{1}{x_n-x_{\sg_{n-1}}}+\frac{x_{\sg_{n-1}}-x_{\sg_{n-2}}}{(x_n-x_{\sg_{n-2}})(x_{\sg_{n-1}}-x_n)}
+\ldots+\frac{x_{\sg_2}-x_{\sg_1}}{(x_n-x_{\sg_1})(x_{\sg_2}-x_n)}+\frac{x_{\sg_1}-y}{(x_n-y)(x_{\sg_1}-x_n)}\\
=\,&\frac{1}{x_n-x_{\sg_1}}+\frac{x_{\sg_1}-y}{(x_n-y)(x_{\sg_1}-x_n)}=\frac{1}{x_n-y},
\eal
\ee
therefore
\be
J_n=\frac{1}{y(x_1-y)\ldots(x_n-y)}\equiv\frac{1}{y(x_1-y)}\cap\cdots\cap\frac{1}{y(x_n-y)}. \labell{eq-5}
\ee
If we evaluate the residue of $J_n$ at $y\!=\!0$, it returns to
\be
J_n(0,X_n)=\int_0\frac{dy}{y}\times\frac{1}{(x_1-y)\ldots(x_n-y)}=\frac{1}{x_1\ldots x_n},
\ee
which is also the completeness relation as all $x_i$'s are trivially greater than zero.
In fact, \eqref{eq-4} and \eqref{eq-5} can be trivially obtained, if we switch to the perspective which considers
$x_1,\ldots,x_n\!<\!y$ and $x_1,\ldots,x_n\!>\!y$ respectively instead. Such an equivalent but much simpler approach
can be further generalized to
\be
\frac{1}{c_1x_1(y-x_1-c_1)}\cap\cdots\cap\frac{1}{c_nx_n(y-x_n-c_n)}
=\frac{y^{n-1}}{c_1\ldots c_n\,x_1\ldots x_n(y-x_1-c_1)\ldots(y-x_n-c_n)},
\ee
as well as
\be
\frac{x_1+c_1}{c_1x_1(x_1+c_1-y)\,y}\cap\cdots\cap\frac{x_n+c_n}{c_nx_n(x_n+c_n-y)\,y}
=\frac{(x_1+c_1)\ldots(x_n+c_n)}{c_1\ldots c_n\,x_1\ldots x_n(x_1+c_1-y)\ldots(x_n+c_n-y)\,y},
\ee
where we have used the expressions in \eqref{eq-3}. A mixed product of these two types is, for example,
\be
\frac{1}{c_1x_1(y-x_1-c_1)}\cap\frac{x_2+c_2}{c_2x_2(x_2+c_2-y)\,y}
=\frac{1}{c_1x_1(y-x_1-c_1)}\times\frac{x_2+c_2}{c_2x_2(x_2+c_2-y)}.
\ee
From these formulas of co-positive products, it is easy to observe that: for $n$ forms that impose positivity conditions
on a number of variables and these conditions involve only one common variable, denoted by $y$ for instance, we have
\be
I_1\cap\cdots\cap I_n=I_1\times\cdots\times I_n\times y^{n-1}, \labell{eq-6}
\ee
which is trivial to prove from the perspective above.
When there are two or more common variables, this simplification is no longer valid
in general. Two such examples are given below:
\be
\frac{y_1}{x_1x_2(y_1-x_1)(y_1-x_2)}\cap\frac{y_2}{x_1x_2(y_2-x_1)(y_2-x_2)}
=\frac{y_1y_2-x_1x_2}{x_1x_2(y_1-x_1)(y_1-x_2)(y_2-x_1)(y_2-x_2)},\,
\ee
as well as
\be
\bal
\!\!
&\frac{y_1^2}{x_1x_2x_3(y_1-x_1)(y_1-x_2)(y_1-x_3)}\cap\frac{y_2^2}{x_1x_2x_3(y_2-x_1)(y_2-x_2)(y_2-x_3)}\\
\!\!
=\,&\frac{y_1^2y_2^2-y_1y_2(x_1x_2+x_1x_3+x_2x_3)+(y_1+y_2)x_1x_2x_3}
{x_1x_2x_3(y_1-x_1)(y_1-x_2)(y_1-x_3)(y_2-x_1)(y_2-x_2)(y_2-x_3)}.
~~~~~~~~~~~~~~~~~~~~~~~~~~~~~~~~~~~~~~~~~~~~
\eal
\ee
It is known that a $d\log$ form can be interpreted as the projective volume, which establishes its relevance with forms.
But in our context, the concept of volume appears more often as the cancelation of spurious poles, for amplitudes
or integrands that are rational functions. This is in fact the key mechanism making summing different terms from
the triangulation of amplituhedron possible.

\newpage
\section{The Trick of Intermediate Variables at 3-loop}
\label{sec3}

Now, we are ready to introduce the trick of intermediate variables to handle the 3 intertwining positivity conditions of
the 4-particle amplituhedron at 3-loop. This is not the final answer that we pursuit, but it divides a difficult problem
into two parts in a pedagogical way, and it is a nice mathematical warmup for the more precise description.
These positivity conditions are
\be
\bal
D_{12}=(x_2-x_1)(z_1-z_2)+(y_2-y_1)(w_1-w_2)>0,\\
D_{13}=(x_3-x_1)(z_1-z_3)+(y_3-y_1)(w_1-w_3)>0,\\
D_{23}=(x_3-x_2)(z_2-z_3)+(y_3-y_2)(w_2-w_3)>0.
\eal
\ee
Without loss of generality, let's work in the ordered subspace $X(123)$ so that $x_1\!<\!x_2\!<\!x_3$.
Then $D_{12}>0$ unambiguously demands
\be
z_1-z_2+\frac{(y_2-y_1)(w_1-w_2)}{x_{21}}>0,
\ee
for instance. Depending on the sign of $(y_2\!-\!y_1)(w_1\!-\!w_2)$, we have
\be
\bal
&(y_2-y_1)(w_1-w_2)>0:~~z_2<z_1+c_{12},~~c_{12}\equiv\frac{(y_2-y_1)(w_1-w_2)}{x_{21}},\\
&(y_2-y_1)(w_1-w_2)<0:~~z_1>z_2+c_{21},~~c_{21}\equiv\frac{-\,(y_2-y_1)(w_1-w_2)}{x_{21}},
\eal
\ee
where $c_{12}$ and $c_{21}$ are defined as the positive \textit{intermediate variables}. The corresponding forms are then
\be
\bal
&Z^-_{12}\equiv\frac{1}{z_1}\(\frac{1}{z_2}-\frac{1}{z_2-z_1-c_{12}}\)=\frac{1}{z_1}\frac{z_1+c_{12}}{z_2(z_1+c_{12}-z_2)},\\
&Z^+_{21}\equiv\frac{1}{z_2}\frac{1}{z_1-z_2-c_{21}},
\eal
\ee
in ordered subspaces $(Y(12)W(21)\!+\!Y(21)W(12))$ and $(Y(12)W(12)\!+\!Y(21)W(21))$ respectively,
and the symbols $Z^+$ and $Z^-$ are related by the completeness relation
\be
Z^+_{21}+Z^-_{21}=\frac{1}{z_1z_2}\equiv I_{12},
\ee
here the identity $I_{12}$ denotes no positivity condition is imposed on $z_1,z_2$.

Therefore, in subspace $X(123)$, for $D_{12},D_{13},D_{23}\!>\!0$ we need to figure out the product
\be
\bal
&\[Z^+_{31}(Y(13)W(13)\!+\!Y(31)W(31))+Z^-_{13}(Y(13)W(31)\!+\!Y(31)W(13))\]\\
\cap&\[Z^+_{32}(Y(23)W(23)\!+\!Y(32)W(32))+Z^-_{23}(Y(23)W(32)\!+\!Y(32)W(23))\]\\
\cap&\[Z^+_{21}(Y(12)W(12)\!+\!Y(21)W(21))+Z^-_{12}(Y(12)W(21)\!+\!Y(21)W(12))\], \labell{eq-7}
\eal
\ee
where the products involving $y$- and $w$-space are easy, so we mainly focus on the products of $Z^\pm$'s.
There are $2^3\!=\!8$ such triple co-positive products, as listed below:
\be
\bal
T_1\equiv Z^+_{31}\cap Z^+_{32}\cap Z^+_{21},~~T_2\equiv Z^+_{31}\cap Z^-_{23}\cap Z^+_{21},\\
T_3\equiv Z^-_{13}\cap Z^+_{32}\cap Z^+_{21},~~T_4\equiv Z^-_{13}\cap Z^-_{23}\cap Z^+_{21},\\
T_5\equiv Z^+_{31}\cap Z^+_{32}\cap Z^-_{12},~~T_6\equiv Z^+_{31}\cap Z^-_{23}\cap Z^-_{12},\\
T_7\equiv Z^-_{13}\cap Z^+_{32}\cap Z^-_{12},~~T_8\equiv Z^-_{13}\cap Z^-_{23}\cap Z^-_{12},
\eal
\ee
and now we will determine them one by one.

For $T_1$, it demands that $z_1$ is greater than both $(z_2\!+\!c_{21})$ and $(z_3\!+\!c_{31})$, so we need to separately
discuss the situations of $z_2\!+\!c_{21}\!>\!z_3\!+\!c_{31}$ and $z_2\!+\!c_{21}\!<\!z_3\!+\!c_{31}$. The extra
complexity is, $z_2$ and $z_3$ are restricted to the subspace of $z_2\!>\!z_3\!+\!c_{32}$.
If $c_{31}\!<\!c_{32}\!+\!c_{21}$, we find $z_2\!+\!c_{21}\!>\!z_3\!+\!c_{32}\!+\!c_{21}\!>\!z_3\!+\!c_{31}$,
so $z_1\!>\!z_2\!+\!c_{21}$ already implies $z_1\!>\!z_3\!+\!c_{31}$. If $c_{31}\!>\!c_{32}\!+\!c_{21}$, and
$z_2\!>\!z_3\!+\!c_{31}\!-\!c_{21}$ which already implies $z_2\!>\!z_3\!+\!c_{32}$, we again
have $z_1\!>\!z_2\!+\!c_{21}$. Finally if $z_3\!+\!c_{32}\!<\!z_2\!<\!z_3\!+\!c_{31}\!-\!c_{21}$, we switch to
$z_1\!>\!z_3\!+\!c_{31}$.

As $c_{31}$ is treated as a positive variable, instead of a rational function of other positive variables
as it actually should be, the discussion above leads to the sum
\be
\bal
T_1=&\(\frac{1}{c_{31}}-\frac{1}{c_{31}-c_{32}-c_{21}}\)\frac{1}{z_3}\frac{1}{z_2-z_3-c_{32}}\frac{1}{z_1-z_2-c_{21}}\\
&+\frac{1}{c_{31}-c_{32}-c_{21}}\frac{1}{z_3}\[\frac{1}{z_2-z_3-c_{31}+c_{21}}\frac{1}{z_1-z_2-c_{21}}\right.\\
&~~~~~~~~~~~~~~~~~~~~~~~~~~~~+\left.\(\frac{1}{z_2-z_3-c_{32}}-\frac{1}{z_2-z_3-c_{31}+c_{21}}\)
\frac{1}{z_1-z_3-c_{31}}\]\\
=\,&\frac{1}{c_{31}}\frac{z_1-z_3}{z_3(z_1-z_2-c_{21})(z_1-z_3-c_{31})(z_2-z_3-c_{32})}, \labell{eq-10}
\eal
\ee
and the $1/c_{31}$ part will be dropped for later convenience. We see this sum wipes off the subspace division of $c_{31}$,
physically it means there is no ``spurious pole''. And if $c_{31}\!=\!0$, $(z_1\!-\!z_3\!-\!c_{31})$ is cancelled,
since this positivity condition becomes redundant as $z_1\!>\!z_2\!+\!c_{21}\!>\!z_3\!+\!c_{32}\!+\!c_{21}$.

Then for $T_2$, $z_2$ and $z_3$ are restricted to the subspace of $z_3\!<\!z_2\!+\!c_{23}$ while other two conditions
remain the same as $T_1$'s. Analogously, we have the following discussion:
\be
\bal
&~c_{21}<c_{31},~~~\,\left\{\begin{array}{c}
z_2>z_3+c_{31}-c_{21},~~~~~~~~~~~~~~~~~~~~~~~~~~~z_1>z_2+c_{21},\\
z_2<z_3+c_{31}-c_{21},~z_3<z_2+c_{23},~~~~~~~~z_1>z_3+c_{31},
\end{array} \right. \\
c_{31}<&~c_{21}<c_{23}+c_{31},~\left\{\begin{array}{c}
~~~~~~~~~~~~~~~~~~~~~\,z_3<z_2+c_{21}-c_{31},~z_1>z_2+c_{21},\\
z_2+c_{21}-c_{31}<z_3<z_2+c_{23},~~~~~~~~\,z_1>z_3+c_{31},
\end{array} \right. \\
&~c_{21}>c_{23}+c_{31},~~~~~~~~~~~~~~~~~~~~~~~~~~z_3<z_2+c_{23},~~~~~~~~\,z_1>z_2+c_{21},
\eal
\ee
note that we have divided the $c_{21}$-space. This leads to
\be
\bal
T_2=&\(\frac{1}{c_{21}}-\frac{1}{c_{21}-c_{31}}\)
\[\frac{1}{z_3}\frac{1}{z_2-z_3-c_{31}+c_{21}}\frac{1}{z_1-z_2-c_{21}}\right.\\
&~~~~~~~~~~~~~~~~~~~~~~~~~~~+\left.\(\frac{1}{z_2}\(\frac{1}{z_3}-\frac{1}{z_3-z_2-c_{23}}\)
-\frac{1}{z_3}\frac{1}{z_2-z_3-c_{31}+c_{21}}\)\frac{1}{z_1-z_3-c_{31}}\]\\
&+\(\frac{1}{c_{21}-c_{31}}-\frac{1}{c_{21}-c_{23}-c_{31}}\)
\frac{1}{z_2}\[\(\frac{1}{z_3}-\frac{1}{z_3-z_2-c_{21}+c_{31}}\)\frac{1}{z_1-z_2-c_{21}}\right.\\
&~~~~~~~~~~~~~~~~~~~~~~~~~~~~~~~~~~~~~~~~~~~~~~~~~+\left.\(\frac{1}{z_3-z_2-c_{21}+c_{31}}-\frac{1}{z_3-z_2-c_{23}}\)
\frac{1}{z_1-z_3-c_{31}}\]\\
&+\frac{1}{c_{21}-c_{23}-c_{31}}\frac{1}{z_2}\(\frac{1}{z_3}-\frac{1}{z_3-z_2-c_{23}}\)\frac{1}{z_1-z_2-c_{21}}\\
=\,&\frac{1}{c_{21}}\frac{z_1(z_2+c_{23})-z_2z_3}{z_2z_3(z_1-z_2-c_{21})(z_1-z_3-c_{31})(z_2+c_{23}-z_3)}. \labell{eq-11}
\eal
\ee
In general, we find that for $c_{ij},c_{jk},c_{ik}$ with respect to $T_1,T_2,T_4,T_5,T_7,T_8$,
it is most convenient to divide the $c_{ik}$-space. While for $c_{ij},c_{jk},c_{ki}$
with respect to $T_3,T_6$, there is no need to divide any of them.

Then for $T_3$, since $z_1\!>\!z_2\!+\!c_{21}\!>\!z_3\!+\!c_{32}\!+\!c_{21}$ already implies $z_3\!<\!z_1\!+\!c_{13}$,
$Z^-_{13}$ becomes redundant, which leads to
\be
T_3=Z^-_{13}\cap Z^+_{32}\cap Z^+_{21}=Z^+_{32}\cap Z^+_{21}=\frac{1}{z_3(z_1-z_2-c_{21})(z_2-z_3-c_{32})}, \labell{eq-12}
\ee
now we see indeed there is no need to consider any $c_{ij}$.

Then for $T_4$, it demands that $z_3$ is less than both $(z_1\!+\!c_{13})$ and $(z_2\!+\!c_{23})$
while $z_1$ and $z_2$ are restricted to the subspace of $z_1\!>\!z_2\!+\!c_{21}$.
Analogously, we have the following discussion:
\be
\bal
&c_{23}<c_{21}+c_{13},~~~~~~~~~~~~~~~~~~\,z_1>z_2+c_{21},~~~~~~~~\,z_3<z_2+c_{23},\\
&c_{23}>c_{21}+c_{13},~\left\{\begin{array}{c}
~~~~~~~~~~~~~~z_1>z_2+c_{23}-c_{13},~z_3<z_2+c_{23},\\
z_2+c_{21}<z_1<z_2+c_{23}-c_{13},~z_3<z_1+c_{13},
\end{array} \right.
\eal
\ee
which leads to
\be
\bal
T_4=&\(\frac{1}{c_{23}}-\frac{1}{c_{23}-c_{21}-c_{13}}\)\frac{1}{z_2}\frac{1}{z_1-z_2-c_{21}}
\(\frac{1}{z_3}-\frac{1}{z_3-z_2-c_{23}}\)\\
&+\frac{1}{c_{23}-c_{21}-c_{13}}\frac{1}{z_2}
\[\frac{1}{z_1-z_2-c_{23}+c_{13}}\(\frac{1}{z_3}-\frac{1}{z_3-z_2-c_{23}}\)\right.\\
&~~~~~~~~~~~~~~~~~~~~~~~~~~~~+\left.\(\frac{1}{z_1-z_2-c_{21}}-\frac{1}{z_1-z_2-c_{23}+c_{13}}\)
\(\frac{1}{z_3}-\frac{1}{z_3-z_1-c_{13}}\)\]\\
=\,&\frac{1}{c_{23}}\frac{(z_1+c_{13})(z_2+c_{23})-z_2z_3}{z_2z_3(z_1-z_2-c_{21})(z_1+c_{13}-z_3)(z_2+c_{23}-z_3)}.
\labell{eq-13}
\eal
\ee
Then for $T_5$, $z_3\!+\!c_{32}\!<\!z_2\!<\!z_1\!+\!c_{12}$ implies that $(z_1\!+\!c_{12})$ must be greater than
$(z_3\!+\!c_{32})$ while $z_1$ and $z_3$ are restricted to the subspace of $z_1\!>\!z_3\!+\!c_{31}$.
Analogously, we have the following discussion:
\be
\bal
&c_{32}<c_{31}+c_{12},~z_1>z_3+c_{31},\\
&c_{32}>c_{31}+c_{12},~z_1>z_3+c_{32}-c_{12},
\eal
\ee
which leads to
\be
\bal
T_5=&\[\(\frac{1}{c_{32}}-\frac{1}{c_{32}-c_{31}-c_{12}}\)\frac{1}{z_1-z_3-c_{31}}
+\frac{1}{c_{32}-c_{31}-c_{12}}\frac{1}{z_1-z_3-c_{32}+c_{12}}\]\\
&\times\frac{1}{z_3}\(\frac{1}{z_2-z_3-c_{32}}-\frac{1}{z_2-z_1-c_{12}}\)\\
=\,&\frac{1}{c_{32}}\frac{(z_1+c_{12})-z_3}{z_3(z_1+c_{12}-z_2)(z_1-z_3-c_{31})(z_2-z_3-c_{32})}. \labell{eq-14}
\eal
\ee
Then for $T_6$, similar to $T_3$, there is no need to consider any $c_{ij}$, the sum is simply
\be
\bal
T_6=\,&\frac{1}{z_1-z_3-c_{31}}\[\frac{1}{z_2}\(\frac{1}{z_3}-\frac{1}{z_3-z_2-c_{23}}\)
-\frac{1}{z_3}\frac{1}{z_2-z_1-c_{12}}\]\\
=\,&\frac{(z_1+c_{12})(z_2+c_{23})-z_2z_3}{z_2z_3(z_1+c_{12}-z_2)(z_2+c_{23}-z_3)(z_1-z_3-c_{31})}. \labell{eq-15}
\eal
\ee
Then for $T_7$, similar to $T_5$, $(z_1\!+\!c_{12})$ must be greater than
$(z_3\!+\!c_{32})$ while $z_1$ and $z_3$ are restricted to the subspace of $z_3\!<\!z_1\!+\!c_{13}$.
Analogously, we have the following discussion:
\be
\bal
&c_{12}<c_{32},\!\!&z_1&>z_3+c_{32}-c_{12},\\
&c_{32}<c_{12}<c_{13}+c_{32},\!\!&z_3&<z_1+c_{12}-c_{32},\\
&c_{12}>c_{13}+c_{32},\!\!&z_3&<z_1+c_{13},
\eal
\ee
which leads to
\be
\bal
T_7=&\[\(\frac{1}{c_{12}}-\frac{1}{c_{12}-c_{32}}\)\frac{1}{z_3}\frac{1}{z_1-z_3-c_{32}+c_{12}}\right.\\
&~~+\(\frac{1}{c_{12}-c_{32}}-\frac{1}{c_{12}-c_{13}-c_{32}}\)
\frac{1}{z_1}\(\frac{1}{z_3}-\frac{1}{z_3-z_1-c_{12}+c_{32}}\)\\
&~~+\left.\frac{1}{c_{12}-c_{13}-c_{32}}\frac{1}{z_1}\(\frac{1}{z_3}-\frac{1}{z_3-z_1-c_{13}}\)\]
\(\frac{1}{z_2-z_3-c_{32}}-\frac{1}{z_2-z_1-c_{12}}\)\\
=\,&\frac{1}{c_{12}}\frac{(z_1+c_{12})(z_1+c_{13})-z_1z_3}{z_1z_3(z_1+c_{12}-z_2)(z_1+c_{13}-z_3)(z_2-z_3-c_{32})}.
\labell{eq-16}
\eal
\ee
{}\\
Finally for $T_8$, similar to $T_4$, $z_3$ is less than both $(z_1\!+\!c_{13})$ and $(z_2\!+\!c_{23})$
while $z_1$ and $z_2$ are restricted to the subspace of $z_2\!<\!z_1\!+\!c_{12}$.
Analogously, we have the following discussion:
\be
\bal
&~c_{13}<c_{23},~~~~~~~~\,\left\{\begin{array}{c}
z_2<z_1+c_{12},~~~z_1<z_2+c_{23}-c_{13},~z_3<z_1+c_{13},\\
~~~~~~~~~~~~~~~~~~~~~\,z_1>z_2+c_{23}-c_{13},~z_3<z_2+c_{23},
\end{array} \right. \\
c_{23}<&~c_{13}<c_{12}+c_{23},~\left\{\begin{array}{c}
z_1+c_{13}-c_{23}<z_2<z_1+c_{12},~~~~~~~~z_3<z_1+c_{13},\\
~~~~~~~~~~~~~~~~~~~~~\,z_2<z_1+c_{13}-c_{23},~z_3<z_2+c_{23},
\end{array} \right. \\
c_{12}+c_{23}<&~c_{13},~~~~~~~~~~~~~~~~~~~~~~~~~~~~~~~~~~~~~~~~~z_2<z_1+c_{12},~~~~~~~~\,z_3<z_2+c_{23},
\eal
\ee
which leads to
\be
\bal
T_8=&\(\frac{1}{c_{13}}-\frac{1}{c_{13}-c_{23}}\)\[\(\frac{1}{z_1}\(\frac{1}{z_2}-\frac{1}{z_2-z_1-c_{12}}\)
-\frac{1}{z_2}\frac{1}{z_1-z_2-c_{23}+c_{13}}\)\(\frac{1}{z_3}-\frac{1}{z_3-z_1-c_{13}}\)\right.\\
&~~~~~~~~~~~~~~~~~~~~~~~~~~~~+\left.\frac{1}{z_2}\frac{1}{z_1-z_2-c_{23}+c_{13}}
\(\frac{1}{z_3}-\frac{1}{z_3-z_2-c_{23}}\)\]\\
&+\(\frac{1}{c_{13}-c_{23}}-\frac{1}{c_{13}-c_{12}-c_{23}}\)\frac{1}{z_1}
\[\(\frac{1}{z_2-z_1-c_{13}+c_{23}}-\frac{1}{z_2-z_1-c_{12}}\)\(\frac{1}{z_3}-\frac{1}{z_3-z_1-c_{13}}\)\right.\\
&~~~~~~~~~~~~~~~~~~~~~~~~~~~~~~~~~~~~~~~~~~~~~~~~~+\left.\(\frac{1}{z_2}-\frac{1}{z_2-z_1-c_{13}+c_{23}}\)
\(\frac{1}{z_3}-\frac{1}{z_3-z_2-c_{23}}\)\]\\
&+\frac{1}{c_{13}-c_{12}-c_{23}}\frac{1}{z_1}\(\frac{1}{z_2}-\frac{1}{z_2-z_1-c_{12}}\)
\(\frac{1}{z_3}-\frac{1}{z_3-z_2-c_{23}}\)\\
=\,&\frac{1}{c_{13}}\frac{(z_1+c_{12})(z_1+c_{13})(z_2+c_{23})-z_1z_2z_3}
{z_1z_2z_3(z_1+c_{12}-z_2)(z_1+c_{13}-z_3)(z_2+c_{23}-z_3)}. \labell{eq-17}
\eal
\ee
Now we have known all eight $T_i$'s. A consistency check via the completeness relation gives
\be
T_2(2\!\leftrightarrow\!3)+T_1=Z^+_{31}\cap Z^+_{21}=Z^+_{31}\times Z^+_{21}\times z_1
=\frac{z_1}{z_2z_3(z_1-z_2-c_{21})(z_1-z_3-c_{31})}, \labell{eq-28}
\ee
where we have used \eqref{eq-6}, similarly we also have (dropping all $1/c_{ij}$ prefactors)
\be
\bal
T_2(1\!\leftrightarrow\!2)+T_7=Z^-_{13}\cap Z^+_{32},\\
T_7(2\!\leftrightarrow\!3)+T_8=Z^-_{13}\cap Z^-_{12},\\
T_5(1\!\leftrightarrow\!2)+T_1=Z^+_{31}\cap Z^+_{32},\\
T_5(2\!\leftrightarrow\!3)+T_4=Z^-_{13}\cap Z^+_{21},\\
T_4(1\!\leftrightarrow\!2)+T_8=Z^-_{13}\cap Z^-_{23},\\
T_6(1\!\leftrightarrow\!2)+T_3=Z^-_{13}\cap Z^+_{32}.
\eal
\ee
These relations, in fact, serve as an equivalent approach to obtain all other $T_i$'s one by one
after we know $T_1$ and $T_3$, following the sequence below:
\be
T_1\to T_2\to T_7\to T_8,~T_1\to T_5\to T_4\to T_8,~T_3\to T_6.
\ee
In addition, we have also observed that from
\be
T_8=\frac{(z_1+c_{12})(z_1+c_{13})(z_2+c_{23})-z_1z_2z_3}{z_1z_2z_3(z_1+c_{12}-z_2)(z_1+c_{13}-z_3)(z_2+c_{23}-z_3)},
\ee
all other $T_i$'s can be obtained via flipping $c_{ij}$ to $-c_{ji}$ in the denominator
with respect to flipping each $Z^-_{ij}$ to $Z^+_{ji}$, as well as setting $c_{ij}$ to zero in the numerator.
Therefore, $T_8$ is named as the \textit{master form}.

There is still another equivalent approach to get the master form which divides the $z$-space instead of
the $c$-space. Defining
\be
\eta_{12}\equiv z_1-z_2+c_{12}>0,~~\eta_{13}\equiv z_1-z_3+c_{13}>0,~~\eta_{23}\equiv z_2-z_3+c_{23}>0,
\ee
we find the sum is then
\be
\bal
T_8=\,&~\,\frac{1}{c_{12}}\(\frac{1}{c_{13}c_{23}}\times Z(321)+\frac{1}{c_{13}\eta_{23}}\times Z(231)
+\frac{1}{\eta_{13}\eta_{23}}\times Z(213)\)\\
&\!\!\!\!+\frac{1}{\eta_{12}}\(\frac{1}{c_{13}c_{23}}\times Z(312)+\frac{1}{c_{13}\eta_{23}}\times Z(132)
+\frac{1}{\eta_{13}\eta_{23}}\times Z(123)\)\\
=\,&\frac{1}{c_{12}c_{13}c_{23}}\frac{(z_1+c_{12})(z_1+c_{13})(z_2+c_{23})-z_1z_2z_3}{z_1z_2z_3\,\eta_{12}\eta_{13}\eta_{23}},
\eal
\ee
as expected. Both ways to get the master form using the completeness relations and
dividing the $z$-space can be generalized beyond 3-loop. Once it is known, we can apply the observation above
to get all $2^{\frac{L(L\!-\!1)}{2}}$ co-positive products of arbitrary $Z^\pm$'s. This observation has not been proved,
but it turns out to be valid at 4-loop. In appendix \ref{app1}, we use the latter way to get the master form at 4-loop
and after that, we check this observation explicitly via two examples, as a mathematical exercise of curiosity.

\newpage
\section{A Naive Sum}
\label{sec4}

Next, we continue to sum the former results over all ordered subspaces, and we find this naive sum which takes the
advantage of intermediate variables ``almost'' reaches the correct answer, as it can reproduce 96 out of
the total 120 monomials in the latter.

To figure out the co-positive products involving $y$- and $w$-space, we define, for instance:
\be
S(12)\equiv Y(12)W(12)+Y(21)W(21),~A(12)\equiv Y(12)W(21)+Y(21)W(12),
\ee
in which the orderings of $y_1,y_2$ and $w_1,w_2$ are the same or opposite respectively. According to \eqref{eq-7},
each $Z^+_{ij}$ is associated with an $S(ij)$, as well as $Z^-_{ij}$ with an $A(ij)$. Then we explicitly figure out
the products of $S$'s and $A$'s with respect to all $T_i$'s as
\be
\bal
T_1:~S_1\equiv S(13)\cap S(23)\cap S(12)=\,&~\,Y(123)W(123)+Y(132)W(132)+Y(213)W(213)\\
&\!\!\!\!+Y(231)W(231)+Y(312)W(312)+Y(321)W(321),\\
T_2:~S_2\equiv S(13)\cap A(23)\cap S(12)=\,&~\,Y(123)W(132)+Y(132)W(123)\\
&\!\!\!\!+Y(231)W(321)+Y(321)W(231),\\
T_3:~S_3\equiv A(13)\cap S(23)\cap S(12)=\,&~\,Y(132)W(312)+Y(312)W(132)\\
&\!\!\!\!+Y(213)W(231)+Y(231)W(213),\\
T_4:~S_4\equiv A(13)\cap A(23)\cap S(12)=\,&~\,Y(123)W(312)+Y(312)W(123)\\
&\!\!\!\!+Y(213)W(321)+Y(321)W(213),\\
T_5:~S_5\equiv S(13)\cap S(23)\cap A(12)=\,&~\,Y(123)W(213)+Y(213)W(123)\\
&\!\!\!\!+Y(312)W(321)+Y(321)W(312),\\
T_6:~S_6\equiv S(13)\cap A(23)\cap A(12)=\,&~\,Y(132)W(213)+Y(213)W(132)\\
&\!\!\!\!+Y(312)W(231)+Y(231)W(312),\\
T_7:~S_7\equiv A(13)\cap S(23)\cap A(12)=\,&~\,Y(123)W(231)+Y(231)W(123)\\
&\!\!\!\!+Y(132)W(321)+Y(321)W(132),\\
T_8:~S_8\equiv A(13)\cap A(23)\cap A(12)=\,&~\,Y(123)W(321)+Y(132)W(231)+Y(213)W(312)\\
&\!\!\!\!+Y(231)W(132)+Y(312)W(213)+Y(321)W(123), \labell{eq-9}
\eal
\ee
note that in particular, above we have used $Y(13)\cap Y(32)\cap Y(21)\!=\!0$ and so forth.
These results are for subspace $X(123)$ only, and we need to consider all other ordered subspaces of $x$, such as
\be
\bal
&X(123):~S_1T_1+S_2T_2+S_3T_3+S_4T_4+S_5T_5+S_6T_6+S_7T_7+S_8T_8,\\
\to\,&X(132):~S_1T_1+S_2T_2+S_3T_5+S_4T_6+S_5T_3+S_6T_4+S_7T_7+S_8T_8,
\eal
\ee
where switching $2\!\leftrightarrow\!3$ for $x,y,z,w$ leads to switching $T_3\!\leftrightarrow\!T_5$ and
$T_4\!\leftrightarrow\!T_6$ as can be easily verified, and the rest pieces are similarly given by
\be
\bal
&X(213):~S_1T_1+S_2T_3+S_3T_2+S_4T_4+S_5T_5+S_6T_7+S_7T_6+S_8T_8,\\
&X(231):~S_1T_1+S_2T_5+S_3T_2+S_4T_6+S_5T_3+S_6T_7+S_7T_4+S_8T_8,\\
&X(312):~S_1T_1+S_2T_3+S_3T_5+S_4T_7+S_5T_2+S_6T_4+S_7T_6+S_8T_8,\\
&X(321):~S_1T_1+S_2T_5+S_3T_3+S_4T_7+S_5T_2+S_6T_6+S_7T_4+S_8T_8.
\eal
\ee
Therefore, summing them over all ordered subspaces of $x,y,w$, we obtain
\be
\textrm{Sum}=(\textrm{Correct answer})-\textrm{Difference},
\ee
where
\be
\bal
&\,(\textrm{Correct answer})\times\textrm{Denominator}\\
=\,&\,(x_2x_3z_1z_2+y_2y_3w_1w_2)D_{13}
+\(x_3^2z_1z_2y_2w_1+x_2x_3z_1^2y_3w_2+x_2z_1y_3^2w_1w_2+x_3z_2y_2y_3w_1^2\)\\
&+(5\textrm{ permutations of 1,2,3}), \labell{eq-8}
\eal
\ee
as well as
\be
\bal
&\,\textrm{Difference}\times\textrm{Denominator}\\
=\,&\,x_2x_3z_1z_2(-\,y_1w_1-y_3w_3)+y_2y_3w_1w_2(-\,x_1z_1-x_3z_3)+(5\textrm{ permutations of 1,2,3}),
\eal
\ee
and we have defined the product of all physical poles as
\be
\textrm{Denominator}\equiv x_1x_2x_3y_1y_2y_3z_1z_2z_3w_1w_2w_3\,D_{12}D_{13}D_{23}. \labell{eq-26}
\ee
Since $D_{ij}$ contains 8 monomials, the correct answer has $(2\!\times\!8\!+\!4)\!\times\!6\!=\!120$ monomials,
and the sum has 96 so their difference has $4\!\times\!6\!=\!24$ monomials. It is important to notice that
terms such as $x_2x_3z_1z_2(-y_1w_1)$ are not dual conformally invariant by themselves, but grouped as
$x_2x_3z_1z_2D_{13}$ they are.

This tentative answer simplified by the trick of intermediate variables captures
more than we expect. Even though it oversimplifies the complexity of $c_{ij}$'s which are functions of $x,y,w$,
it still gives most parts of the correct answer. If we manually heal the dual conformal invariance,
it is then correct. Remarkably, even if it does not give the full numerator, it can wipe off
the subspace division of all positive variables, which frees it from spurious poles.
After we refine the calculation in order to reach the correct answer, we will return to discuss the diagrammatic
interpretation of \eqref{eq-8}.

\newpage
\section{Refined Co-positive Products}
\label{sec5}

To precisely describe the 4-particle amplituhedron at 3-loop,
we need to further refine co-positive products for each ordered subspace of $y$ and $w$,
based on the former discussions using intermediate variables. These seemingly lengthy results can be nicely rearranged
in order to manifest its simple mathematical structure, namely the Mondrian diagrammatic interpretation.

From the previous setting we know that, for each ordered subspace of $x$, there are eight $T_i$'s, namely the
co-positive products in terms of intermediate variables. From \eqref{eq-9},
each of $T_1,T_8$ corresponds to six ordered subspaces of $y$ and $w$,
while each of $T_2,T_3,T_4,T_5,T_7,T_8$ corresponds to four, so that in total their number is $6\!\times\!6\!=\!36$
as expected. If we abandon intermediate variables, in principle we have to figure out 36 co-positive products instead of 8,
as elaborated in the following.

For $T_1$, the six different ordered subspaces lead to six different sets of $c_{ij}$'s. First for $Y(123)W(123)$,
the condition $c_{31}\!>\!c_{32}\!+\!c_{21}$ is now replaced by
\be
\frac{(y_{32}+y_{21})(w_{32}+w_{21})}{x_{32}+x_{21}}>\frac{y_{32}w_{32}}{x_{32}}+\frac{y_{21}w_{21}}{x_{21}}
\Longrightarrow\(\frac{y_{32}}{y_{21}}-\frac{x_{32}}{x_{21}}\)\(\frac{w_{32}}{w_{21}}-\frac{x_{32}}{x_{21}}\)<0,
\ee
where $x_{32},x_{21},y_{32},y_{21},w_{32},w_{21}$ are positive variables in this subspace
(as usual, we first work in $X(123)$). The corresponding form is
\be
\bal
&\frac{1}{y_{21}}\frac{1}{y_{32}-y_{21}x_{32}/x_{21}}
\frac{1}{w_{32}}\frac{1}{w_{21}-w_{32}x_{21}/x_{32}}+(y\leftrightarrow w)\\
=\,&\frac{x_{21}x_{32}(y_{21}w_{32}+y_{32}w_{21})}
{y_{21}y_{32}\,w_{21}w_{32}(x_{32}y_{21}-x_{21}y_{32})(-\,x_{32}w_{21}+x_{21}w_{32})}, \labell{eq-18}
\eal
\ee
then using the completeness relation, the form of $c_{31}\!<\!c_{32}\!+\!c_{21}$ is
\be
\bal
&\frac{1}{y_{21}y_{32}\,w_{21}w_{32}}-\frac{x_{21}x_{32}(y_{21}w_{32}+y_{32}w_{21})}
{y_{21}y_{32}\,w_{21}w_{32}(x_{32}y_{21}-x_{21}y_{32})(-\,x_{32}w_{21}+x_{21}w_{32})}\\
=\,&\frac{x_{32}^2y_{21}w_{21}+x_{21}^2y_{32}w_{32}}
{y_{21}y_{32}\,w_{21}w_{32}(x_{32}y_{21}-x_{21}y_{32})(x_{32}w_{21}-x_{21}w_{32})}.
\eal
\ee
Making relevant replacements in \eqref{eq-10} gives the form
\be
T_{1,\,Y(123)W(123)}=\frac{1}{D_{12}D_{13}D_{23}}\,Y(123)W(123)\,\frac{D_{13}+y_{21}w_{32}+y_{32}w_{21}}{x_1z_3},
\labell{eq-19}
\ee
analogously, for $Y(321)W(321)$ it becomes
\be
T_{1,\,Y(321)W(321)}=\frac{1}{D_{12}D_{13}D_{23}}\,Y(321)W(321)\,\frac{D_{13}+y_{12}w_{23}+y_{23}w_{12}}{x_1z_3}.
\ee
For convenience, we will drop the prefactors $Y(\ldots)W(\ldots)/D_{12}D_{13}D_{23}$ below.

Next for $Y(132)W(132)$, we trivially have $c_{31}\!<\!c_{21}$ since
\be
\frac{y_{31}w_{31}}{x_{32}+x_{21}}<\frac{(y_{23}+y_{31})(w_{23}+w_{31})}{x_{21}}
\ee
always holds, which belongs to the situation of $c_{31}\!<\!c_{32}\!+\!c_{21}$. Therefore the sum in \eqref{eq-10}
trivially has one term only, which gives
\be
T_{1,\,Y(132)W(132)}=\frac{D_{13}}{x_1z_3},
\ee
analogously, for $Y(231)W(231)$ it becomes
\be
T_{1,\,Y(231)W(231)}=\frac{D_{13}}{x_1z_3}.
\ee
Last for $Y(213)W(213)$, we trivially have $c_{31}\!<\!c_{32}$ since
\be
\frac{y_{31}w_{31}}{x_{32}+x_{21}}<\frac{(y_{31}+y_{12})(w_{31}+w_{12})}{x_{32}}
\ee
always holds, which belongs to the situation of $c_{31}\!<\!c_{32}\!+\!c_{21}$. Therefore the sum in \eqref{eq-10}
trivially has one term only, which gives
\be
T_{1,\,Y(213)W(213)}=\frac{D_{13}}{x_1z_3},
\ee
as well as for $Y(312)W(312)$,
\be
T_{1,\,Y(312)W(312)}=\frac{D_{13}}{x_1z_3}.
\ee
\\
Then for $T_2$ in $Y(123)W(132)$, the condition $c_{21}\!<\!c_{31}$ is replaced by
\be
\frac{y_{21}(w_{23}+w_{31})}{x_{21}}<\frac{(y_{32}+y_{21})w_{31}}{x_{32}+x_{21}}\Longrightarrow
\frac{y_{32}}{y_{21}}>\frac{x_{32}}{x_{21}}+\frac{w_{23}}{w_{31}}+\frac{x_{32}}{x_{21}}\frac{w_{23}}{w_{31}}\equiv\ap,
\ee
and the corresponding form is
\be
\frac{1}{y_{21}}\frac{1}{y_{32}-y_{21}\ap},
\ee
similarly, $c_{21}\!>\!c_{23}\!+\!c_{31}$ is replaced by
\be
\frac{y_{21}(w_{23}+w_{31})}{x_{21}}>\frac{y_{32}w_{23}}{x_{32}}+\frac{(y_{32}+y_{21})w_{31}}{x_{32}+x_{21}}
\Longrightarrow\frac{y_{32}}{y_{21}}<\frac{x_{32}}{x_{21}},
\ee
and the corresponding form is replaced by
\be
\frac{1}{y_{32}}\frac{1}{y_{21}-y_{32}x_{21}/x_{32}},
\ee
hence the form of $c_{31}\!<\!c_{21}\!<\!c_{23}\!+\!c_{31}$ is
\be
\frac{1}{y_{21}y_{32}}-\frac{1}{y_{21}}\frac{1}{y_{32}-y_{21}\ap}-\frac{1}{y_{32}}\frac{1}{y_{21}-y_{32}x_{21}/x_{32}}.
\ee
Making relevant replacements in \eqref{eq-11} gives
\be
T_{2,\,Y(123)W(132)}=\frac{x_{32}+x_{21}}{x_1z_2z_3}
\[\(z_1-\frac{y_{21}w_{31}}{x_{32}+x_{21}}\)\(z_2+\frac{y_{32}w_{23}}{x_{32}}\)-z_2z_3\],
\ee
analogously, for $Y(321)W(231)$ it becomes
\be
T_{2,\,Y(321)W(231)}=\frac{x_{32}+x_{21}}{x_1z_2z_3}
\[\(z_1-\frac{y_{12}w_{13}}{x_{32}+x_{21}}\)\(z_2+\frac{y_{23}w_{32}}{x_{32}}\)-z_2z_3\].
\ee
Switching $y\!\leftrightarrow\!w$, for $Y(132)W(123)$ and $Y(231)W(321)$ we obtain
\be
\bal
T_{2,\,Y(132)W(123)}=\frac{x_{32}+x_{21}}{x_1z_2z_3}
\[\(z_1-\frac{y_{31}w_{21}}{x_{32}+x_{21}}\)\(z_2+\frac{y_{23}w_{32}}{x_{32}}\)-z_2z_3\],\\
T_{2,\,Y(231)W(321)}=\frac{x_{32}+x_{21}}{x_1z_2z_3}
\[\(z_1-\frac{y_{13}w_{12}}{x_{32}+x_{21}}\)\(z_2+\frac{y_{32}w_{23}}{x_{32}}\)-z_2z_3\].
\eal
\ee
\\
Then for $T_3$ in $Y(132)W(312)$, $Y(312)W(132)$, $Y(213)W(231)$ and $Y(231)W(213)$, there is no difference in
the discussions as we do not divide any $c_{ij}$. Immediately, \eqref{eq-12} gives
\be
\bal
T_{3,\,Y(132)W(312)}=\frac{D_{13}}{x_1z_3},~~T_{3,\,Y(231)W(213)}=\frac{D_{13}}{x_1z_3},\\
T_{3,\,Y(312)W(132)}=\frac{D_{13}}{x_1z_3},~~T_{3,\,Y(213)W(231)}=\frac{D_{13}}{x_1z_3},
\eal
\ee
note that we must separately write these identical results, since they have different hidden prefactors.
\\ \\
Then for $T_4$ in $Y(123)W(312)$, the condition $c_{23}\!<\!c_{21}\!+\!c_{13}$ is replaced by
\be
\frac{y_{32}(w_{21}+w_{13})}{x_{32}}<\frac{y_{21}w_{21}}{x_{21}}+\frac{(y_{32}+y_{21})w_{13}}{x_{32}+x_{21}}
\Longrightarrow\frac{y_{32}}{y_{21}}<\frac{x_{32}}{x_{21}},
\ee
and the corresponding form is
\be
\frac{1}{y_{32}}\frac{1}{y_{21}-y_{32}x_{21}/x_{32}},
\ee
similarly, the form of $c_{23}\!>\!c_{21}\!+\!c_{13}$ is
\be
\frac{1}{y_{21}}\frac{1}{y_{32}-y_{21}x_{32}/x_{21}}.
\ee
Making relevant replacements in \eqref{eq-13} gives
\be
\!\!\!\!\!\!\!\!\!
T_{4,\,Y(123)W(312)}=\frac{x_{32}+x_{21}}{x_1z_2z_3}
\[\(z_1+\frac{(y_{32}+y_{21})w_{13}}{x_{32}+x_{21}}\)\(z_2+\frac{y_{32}(w_{21}+w_{13})}{x_{32}}\)
-\(z_2+\frac{y_{32}w_{13}}{x_{32}+x_{21}}\)z_3\],
\ee
analogously, for $Y(321)W(213)$ it becomes
\be
\!\!\!\!\!\!\!\!\!
T_{4,\,Y(321)W(213)}=\frac{x_{32}+x_{21}}{x_1z_2z_3}
\[\(z_1+\frac{(y_{12}+y_{23})w_{31}}{x_{32}+x_{21}}\)\(z_2+\frac{y_{23}(w_{31}+w_{12})}{x_{32}}\)
-\(z_2+\frac{y_{23}w_{31}}{x_{32}+x_{21}}\)z_3\].
\ee
Switching $y\!\leftrightarrow\!w$, for $Y(312)W(123)$ and $Y(213)W(321)$ we obtain
\be
\bal
\!\!\!\!\!\!\!\!\!\!\!\!\!\!\!\!\!
T_{4,\,Y(312)W(123)}=\frac{x_{32}+x_{21}}{x_1z_2z_3}
\[\(z_1+\frac{y_{13}(w_{32}+w_{21})}{x_{32}+x_{21}}\)\(z_2+\frac{(y_{21}+y_{13})w_{32}}{x_{32}}\)
-\(z_2+\frac{y_{13}w_{32}}{x_{32}+x_{21}}\)z_3\],\\
\!\!\!\!\!\!\!\!\!\!\!\!\!\!\!\!\!
T_{4,\,Y(213)W(321)}=\frac{x_{32}+x_{21}}{x_1z_2z_3}
\[\(z_1+\frac{y_{31}(w_{12}+w_{23})}{x_{32}+x_{21}}\)\(z_2+\frac{(y_{31}+y_{12})w_{23}}{x_{32}}\)
-\(z_2+\frac{y_{31}w_{23}}{x_{32}+x_{21}}\)z_3\].
\eal
\ee
\\
Then for $T_5$ in $Y(123)W(213)$, the condition $c_{32}\!<\!c_{31}\!+\!c_{12}$ is replaced by
\be
\frac{y_{32}(w_{31}+w_{12})}{x_{32}}<\frac{(y_{32}+y_{21})w_{31}}{x_{32}+x_{21}}+\frac{y_{21}w_{12}}{x_{21}}
\Longrightarrow\frac{y_{32}}{y_{21}}<\frac{x_{32}}{x_{21}},
\ee
and  the corresponding form is
\be
\frac{1}{y_{32}}\frac{1}{y_{21}-y_{32}x_{21}/x_{32}}.
\ee
Making relevant replacements in \eqref{eq-14} gives
\be
T_{5,\,Y(123)W(213)}=\frac{x_{32}+x_{21}}{x_1z_3}\(z_1-z_3-\frac{y_{32}w_{31}}{x_{32}+x_{21}}+\frac{y_{21}w_{12}}{x_{21}}\),
\ee
analogously, for $Y(321)W(312)$ it becomes
\be
T_{5,\,Y(321)W(312)}=\frac{x_{32}+x_{21}}{x_1z_3}\(z_1-z_3-\frac{y_{23}w_{13}}{x_{32}+x_{21}}+\frac{y_{12}w_{21}}{x_{21}}\).
\ee
Switching $y\!\leftrightarrow\!w$, for $Y(213)W(123)$ and $Y(312)W(321)$ we obtain
\be
\bal
T_{5,\,Y(213)W(123)}=\frac{x_{32}+x_{21}}{x_1z_3}\(z_1-z_3-\frac{y_{31}w_{32}}{x_{32}+x_{21}}+\frac{y_{12}w_{21}}{x_{21}}\),\\
T_{5,\,Y(312)W(321)}=\frac{x_{32}+x_{21}}{x_1z_3}\(z_1-z_3-\frac{y_{13}w_{23}}{x_{32}+x_{21}}+\frac{y_{21}w_{12}}{x_{21}}\).
\eal
\ee
\\
Then for $T_6$ in $Y(132)W(213)$, $Y(213)W(132)$, $Y(312)W(231)$ and $Y(231)W(312)$, there is no difference in
the discussions, similar to the case of $T_3$. But the different $c_{ij}$'s now matter, as \eqref{eq-15} gives
\be
\bal
T_{6,\,Y(132)W(213)}=\frac{x_{32}+x_{21}}{x_1z_2z_3}
\[\(z_1+\frac{(y_{23}+y_{31})w_{12}}{x_{21}}\)\(z_2+\frac{y_{23}(w_{31}+w_{12})}{x_{32}}\)-z_2z_3\],\\
T_{6,\,Y(231)W(312)}=\frac{x_{32}+x_{21}}{x_1z_2z_3}
\[\(z_1+\frac{(y_{13}+y_{32})w_{21}}{x_{21}}\)\(z_2+\frac{y_{32}(w_{21}+w_{13})}{x_{32}}\)-z_2z_3\],
\eal
\ee
switching $y\!\leftrightarrow\!w$, we obtain
\be
\bal
T_{6,\,Y(213)W(132)}=\frac{x_{32}+x_{21}}{x_1z_2z_3}
\[\(z_1+\frac{y_{12}(w_{23}+w_{31})}{x_{21}}\)\(z_2+\frac{(y_{31}+y_{12})w_{23}}{x_{32}}\)-z_2z_3\],\\
T_{6,\,Y(312)W(231)}=\frac{x_{32}+x_{21}}{x_1z_2z_3}
\[\(z_1+\frac{y_{21}(w_{13}+w_{32})}{x_{21}}\)\(z_2+\frac{(y_{21}+y_{13})w_{32}}{x_{32}}\)-z_2z_3\].
\eal
\ee
\\
Then for $T_7$ in $Y(123)W(231)$, the condition $c_{12}\!<\!c_{32}$ is replaced by
\be
\frac{y_{21}(w_{13}+w_{32})}{x_{21}}<\frac{y_{32}w_{32}}{x_{32}}
\Longrightarrow\frac{y_{32}}{y_{21}}>\frac{x_{32}}{x_{21}}\(1+\frac{w_{13}}{w_{32}}\)\equiv\bt,
\ee
and the corresponding form is
\be
\frac{1}{y_{21}}\frac{1}{y_{32}-y_{21}\bt},
\ee
similarly, $c_{12}\!>\!c_{13}\!+\!c_{32}$ is replaced by
\be
\frac{y_{21}(w_{13}+w_{32})}{x_{21}}>\frac{(y_{32}+y_{21})w_{13}}{x_{32}+x_{21}}+\frac{y_{32}w_{32}}{x_{32}}
\Longrightarrow\frac{y_{32}}{y_{21}}<\frac{x_{32}}{x_{21}},
\ee
and the corresponding form is
\be
\frac{1}{y_{32}}\frac{1}{y_{21}-y_{32}x_{21}/x_{32}}.
\ee
Making relevant replacements in \eqref{eq-16} gives
\be
\!\!\!\!\!\!\!\!\!
T_{7,\,Y(123)W(231)}=\frac{x_{32}+x_{21}}{x_1z_1z_3}
\[\(z_1+\frac{y_{21}(w_{13}+w_{32})}{x_{21}}\)\(z_1+\frac{(y_{32}+y_{21})w_{13}}{x_{32}+x_{21}}\)
-\(z_1+\frac{y_{21}w_{13}}{x_{32}+x_{21}}\)z_3\],
\ee
analogously, for $Y(321)W(132)$ it becomes
\be
\!\!\!\!\!\!\!\!\!
T_{7,\,Y(321)W(132)}=\frac{x_{32}+x_{21}}{x_1z_1z_3}
\[\(z_1+\frac{y_{12}(w_{23}+w_{31})}{x_{21}}\)\(z_1+\frac{(y_{12}+y_{23})w_{31}}{x_{32}+x_{21}}\)
-\(z_1+\frac{y_{12}w_{31}}{x_{32}+x_{21}}\)z_3\].
\ee
Switching $y\!\leftrightarrow\!w$, for $Y(231)W(123)$ and $Y(132)W(321)$ we obtain
\be
\bal
\!\!\!\!\!\!\!\!\!\!\!\!\!\!\!\!\!
T_{7,\,Y(231)W(123)}=\frac{x_{32}+x_{21}}{x_1z_1z_3}
\[\(z_1+\frac{(y_{13}+y_{32})w_{21}}{x_{21}}\)\(z_1+\frac{y_{13}(w_{32}+w_{21})}{x_{32}+x_{21}}\)
-\(z_1+\frac{y_{13}w_{21}}{x_{32}+x_{21}}\)z_3\],\\
\!\!\!\!\!\!\!\!\!\!\!\!\!\!\!\!\!
T_{7,\,Y(132)W(321)}=\frac{x_{32}+x_{21}}{x_1z_1z_3}
\[\(z_1+\frac{(y_{23}+y_{31})w_{12}}{x_{21}}\)\(z_1+\frac{y_{31}(w_{12}+w_{23})}{x_{32}+x_{21}}\)
-\(z_1+\frac{y_{31}w_{12}}{x_{32}+x_{21}}\)z_3\].
\eal
\ee
\\
Finally for $T_8$ in $Y(123)W(321)$, the condition $c_{13}\!<\!c_{23}$ is replaced by
\be
\frac{(y_{32}+y_{21})(w_{12}+w_{23})}{x_{32}+x_{21}}<\frac{y_{32}w_{23}}{x_{32}}\Longrightarrow
\frac{x_{21}}{x_{32}}>\frac{y_{21}}{y_{32}}+\frac{w_{12}}{w_{23}}+\frac{y_{21}}{y_{32}}\frac{w_{12}}{w_{23}}\equiv\gm,
\ee
and the corresponding form is
\be
\frac{1}{x_{32}}\frac{1}{x_{21}-x_{32}\gm},
\ee
similarly, $c_{13}\!>\!c_{12}\!+\!c_{23}$ is replaced by
\be
\frac{(y_{32}+y_{21})(w_{12}+w_{23})}{x_{32}+x_{21}}>\frac{y_{21}w_{12}}{x_{21}}+\frac{y_{32}w_{23}}{x_{32}}
\Longrightarrow\(\frac{y_{32}}{y_{21}}-\frac{x_{32}}{x_{21}}\)\(\frac{w_{23}}{w_{12}}-\frac{x_{32}}{x_{21}}\)<0,
\ee
by trivially adjusting the prefactors of \eqref{eq-18}, we get the corresponding form
\be
\frac{(y_{21}w_{32}+y_{32}w_{21})}{(x_{32}y_{21}-x_{21}y_{32})(-\,x_{32}w_{21}+x_{21}w_{32})}.
\ee
Making relevant replacements in \eqref{eq-17} gives
\be
\bal
T_{8,\,Y(123)W(321)}=\frac{x_{32}+x_{21}}{x_1z_1z_2z_3}
\[\(z_2+\frac{y_{32}w_{23}}{x_{32}}\)\(z_1+\frac{y_{21}w_{12}}{x_{21}}\)
\(z_1+\frac{(y_{32}+y_{21})(w_{12}+w_{23})}{x_{32}+x_{21}}\)\right.\\
\left.-\(\frac{y_{21}w_{12}}{x_{32}+x_{21}}\(z_2+\frac{y_{32}w_{23}}{x_{21}}\)
+\(z_2+\frac{y_{32}w_{23}}{x_{32}+x_{21}}\)z_1\)z_3\],~~~
\eal
\ee
analogously, for $Y(321)W(123)$ it becomes
\be
\bal
T_{8,\,Y(321)W(123)}=\frac{x_{32}+x_{21}}{x_1z_1z_2z_3}
\[\(z_2+\frac{y_{23}w_{32}}{x_{32}}\)\(z_1+\frac{y_{12}w_{21}}{x_{21}}\)
\(z_1+\frac{(y_{12}+y_{23})(w_{32}+w_{21})}{x_{32}+x_{21}}\)\right.\\
\left.-\(\frac{y_{12}w_{21}}{x_{32}+x_{21}}\(z_2+\frac{y_{23}w_{32}}{x_{21}}\)
+\(z_2+\frac{y_{23}w_{32}}{x_{32}+x_{21}}\)z_1\)z_3\].~~~\,
\eal
\ee
Next for $Y(132)W(231)$, we trivially have $c_{13}\!<\!c_{12}$ since
\be
\frac{y_{31}w_{13}}{x_{32}+x_{21}}<\frac{(y_{23}+y_{31})(w_{13}+w_{32})}{x_{21}}
\ee
always holds, which forbids the situation of $c_{13}\!>\!c_{12}\!+\!c_{23}$ so we only need to discuss whether
$c_{13}$ is greater than $c_{23}$. The condition $c_{13}\!<\!c_{23}$ is replaced by
\be
\frac{y_{31}w_{13}}{x_{32}+x_{21}}<\frac{y_{23}w_{32}}{x_{32}}\Longrightarrow
\frac{y_{31}}{y_{23}}<\(1+\frac{x_{21}}{x_{32}}\)\frac{w_{32}}{w_{13}}\equiv\de,
\ee
and the corresponding form is
\be
\frac{1}{y_{31}}\frac{1}{y_{23}-y_{31}/\de}.
\ee
Making relevant replacements in \eqref{eq-17} gives
\be
\!\!\!\!\!\!\!\!\!\!\!\!\!\!\!
T_{8,\,Y(132)W(231)}=\frac{x_{32}+x_{21}}{x_1z_1z_2z_3}\(z_1+\frac{y_{31}w_{13}}{x_{32}+x_{21}}\)
\[\(z_1+\frac{(y_{23}+y_{31})(w_{13}+w_{32})}{x_{21}}\)\(z_2+\frac{y_{23}w_{32}}{x_{32}}\)-z_2z_3\],
\ee
analogously, for $Y(231)W(132)$ it becomes
\be
\!\!\!\!\!\!\!\!\!\!\!\!\!\!\!
T_{8,\,Y(231)W(132)}=\frac{x_{32}+x_{21}}{x_1z_1z_2z_3}\(z_1+\frac{y_{13}w_{31}}{x_{32}+x_{21}}\)
\[\(z_1+\frac{(y_{13}+y_{32})(w_{23}+w_{31})}{x_{21}}\)\(z_2+\frac{y_{32}w_{23}}{x_{32}}\)-z_2z_3\].
\ee
Last for $Y(213)W(312)$, we trivially have $c_{13}\!<\!c_{23}$ since
\be
\frac{y_{31}w_{13}}{x_{32}+x_{21}}<\frac{(y_{31}+y_{12})(w_{21}+w_{13})}{x_{32}}
\ee
always holds. Therefore the sum in \eqref{eq-17} trivially has one term only, which gives
\be
\bal
T_{8,\,Y(213)W(312)}=\frac{x_{32}+x_{21}}{x_1z_1z_2z_3}
\[\(z_1+\frac{y_{12}w_{21}}{x_{21}}\)\(z_1+\frac{y_{31}w_{13}}{x_{32}+x_{21}}\)
\(z_2+\frac{(y_{31}+y_{12})(w_{21}+w_{13})}{x_{32}}\)\right.\\
\left.-\,z_3\(\frac{y_{31}w_{13}}{x_{32}+x_{21}}\(z_1+\frac{y_{12}w_{21}}{x_{21}}\)+z_1z_2\)\],
~~~~~~~~~~~~~~~~~~~~~~~~~~~~
\eal
\ee
analogously, for $Y(312)W(213)$ it becomes
\be
\bal
T_{8,\,Y(312)W(213)}=\frac{x_{32}+x_{21}}{x_1z_1z_2z_3}
\[\(z_1+\frac{y_{21}w_{12}}{x_{21}}\)\(z_1+\frac{y_{13}w_{31}}{x_{32}+x_{21}}\)
\(z_2+\frac{(y_{21}+y_{13})(w_{31}+w_{12})}{x_{32}}\)\right.\\
\left.-\,z_3\(\frac{y_{13}w_{31}}{x_{32}+x_{21}}\(z_1+\frac{y_{21}w_{12}}{x_{21}}\)+z_1z_2\)\].
~~~~~~~~~~~~~~~~~~~~~~~~~~~~\,
\eal
\ee
This finishes the derivation of all 36 co-positive products but most of them can be trivially related to each other,
as there are only 11 independent cases to be analyzed.

\newpage
\section{The Correct Sum and its Mondrian Diagrammatic Interpretation}
\label{sec6}

Collecting the 36 co-positive products for all ordered subspaces of $y$ and $w$, we can continue to sum these results,
and this time, it indeed reaches the correct answer. Instead of a brute-force summation, for each piece
we delicately separate the contributing and the spurious parts.
The former manifest the Mondrian diagrammatic interpretation with which they nicely sum to \eqref{eq-8},
while the latter sum to zero at the end.

\begin{figure}
\begin{center}
\includegraphics[width=0.75\textwidth]{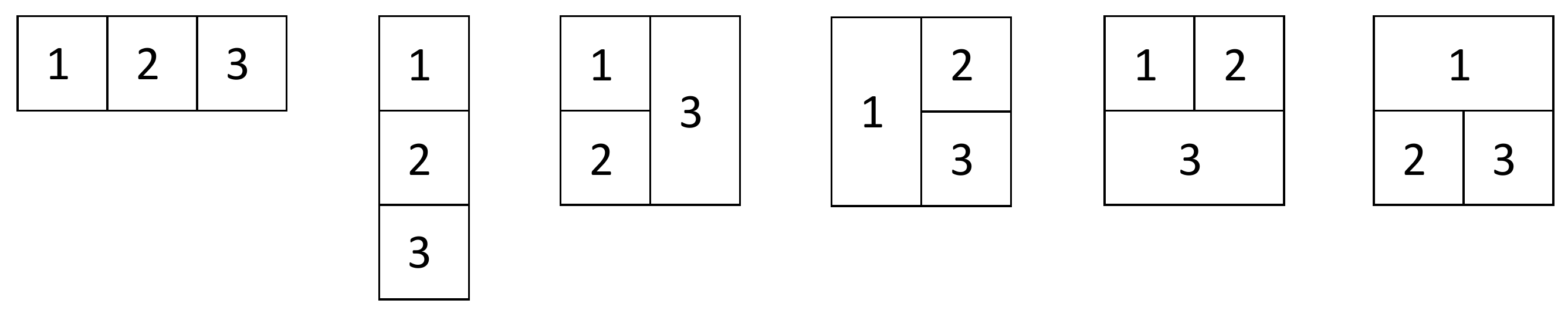}
\caption{Legal seed diagrams in $X(123)Y(123)$: ladders and tennis courts.} \label{fig-2}
\end{center}
\end{figure}

$\!\!\!\!\!\!\!\!\!\!\!\!\!$
The first example is $Y(123)W(123)$, of which the form \eqref{eq-19} can be rewritten as
\be
\bal
T_{1,\,Y(123)W(123)}=\,&\frac{1}{D_{12}D_{13}D_{23}}\,Y(123)W(123)\,\frac{D_{13}+y_{21}w_{32}+y_{32}w_{21}}{x_1z_3}\\
=\,&\frac{1}{D_{12}D_{13}D_{23}\,z_1z_2z_3}\,X(123)Y(123)W(123)\times x_{21}x_{32}z_1z_2(D_{13}+y_{21}w_{32}+y_{32}w_{21})\\
=\,&(\textrm{prefactors})\times(x_{21}x_{32}z_1z_2D_{13}+x_{21}x_{32}z_1z_2(y_{21}w_{32}+y_{32}w_{21})),
\eal
\ee
where again we will drop the prefactors that simply encode its information of ordered subspaces as well as physical poles.
The first term above denotes the seed diagram which pictorially is a horizontal ladder,
the first diagram given in figure \ref{fig-2}. According to the contact rules conceived in the introduction,
since boxes $1,2$ have a horizontal contact and so do boxes $2,3$, we can trivially read off the factor
$x_{21}x_{32}z_1z_2D_{13}$ from that ladder diagram.
In fact, this factor originates from $x_{21}x_{32}z_{12}z_{23}D_{13}$ in the ordered subspace $Z(321)$,
before we sum over all subspaces of $z$ that admit it. As we have known $Y(123)W(123)$ forbids any vertical contact
of boxes (or loops), so we only have a horizontal ladder for this subspace, while the rest terms are spurious.
For later convenience, we can define
\be
\bal
&T_{1,\,Y(123)W(123)}=(100000)+S_{1,\,Y(123)W(123)},\\
&S_{1,\,Y(123)W(123)}\equiv x_{21}x_{32}z_1z_2(y_{21}w_{32}+y_{32}w_{21}),
\eal
\ee
where the symbol $(100000)$ denotes which ones are present for $W(123)$ among the six legal seed diagrams for $X(123)Y(123)$
in figure \ref{fig-2}, as the latter cover two distinct topologies of various orientations at 3-loop.
The numbers filled in the boxes above are uniquely determined by the orderings $X(123)Y(123)$, and if the ordering of $w$
conflicts with that of $y$, the relevant diagram is excluded.

Analogously, for $W(132)$ we have
\be
\bal
&T_{2,\,Y(123)W(132)}=x_{21}x_{32}z_1z_2D_{13}+x_{21}(x_{32}+x_{21})z_1^2y_{32}w_{23}+S_{2,\,Y(123)W(132)},\\
&~~~~~~~~~~~~~~~~~~~=(100100)+S_{2,\,Y(123)W(132)},\\
&S_{2,\,Y(123)W(132)}=x_{21}z_1(x_{32}z_2-y_{21}w_{23})y_{32}w_{31},
\eal
\ee
where $(100100)$ denotes the first and fourth diagrams in figure \ref{fig-2} are present. Note that, $(x_{32}\!+\!x_{21})$
in the first line is nothing but $(x_3\!-\!x_1)$, which means a horizontal contact between boxes $1,3$.

The results for the rest four orderings of $w$ and $Y(123)$ are given by
\be
\bal
&T_{5,\,Y(123)W(213)}=x_{21}x_{32}z_1z_2D_{13}+(x_{32}+x_{21})x_{32}z_1z_2y_{21}w_{12}+S_{5,\,Y(123)W(213)},\\
&~~~~~~~~~~~~~~~~~~~=(101000)+S_{5,\,Y(123)W(213)},\\
&S_{5,\,Y(123)W(213)}=x_{21}x_{32}z_1z_2y_{21}w_{31}
\eal
\ee
for $W(213)$, and
\be
\bal
&T_{7,\,Y(123)W(231)}=x_{21}x_{32}z_1z_2D_{13}+(x_{32}+x_{21})x_{32}z_1z_2y_{21}(w_{13}+w_{32})
+x_{32}z_2y_{21}(y_{32}+y_{21})(w_{13}+w_{32})w_{13}\\
&~~~~~~~~~~~~~~~~~~~~~~+S_{7,\,Y(123)W(231)},\\
&~~~~~~~~~~~~~~~~~~~=(101001)+S_{7,\,Y(123)W(231)},\\
&S_{7,\,Y(123)W(231)}=-\,x_{21}x_{32}z_2z_3y_{21}w_{13}
\eal
\ee
for $W(231)$, and
\be
\bal
&T_{4,\,Y(123)W(312)}=x_{21}x_{32}z_1z_2D_{13}
+x_{21}(x_{32}+x_{21})z_1^2y_{32}(w_{21}+w_{13})+x_{21}z_1(y_{32}+y_{21})y_{32}w_{13}(w_{21}+w_{13})\\
&~~~~~~~~~~~~~~~~~~~~~~+S_{4,\,Y(123)W(312)},\\
&~~~~~~~~~~~~~~~~~~~=(100110)+S_{4,\,Y(123)W(312)},\\
&S_{4,\,Y(123)W(312)}=-\,x_{21}x_{32}z_1z_3y_{32}w_{13}
\eal
\ee
for $W(312)$, and
\be
\bal
&T_{8,\,Y(123)W(321)}=x_{21}x_{32}z_1z_2D_{13}+y_{21}y_{32}w_{12}w_{23}D_{13}\\
&~~~~~~~~~~~~~~~~~~~~~~+(x_{32}+x_{21})x_{32}z_1z_2y_{21}w_{12}+x_{21}(x_{32}+x_{21})z_1^2y_{32}w_{23}\\
&~~~~~~~~~~~~~~~~~~~~~~+x_{21}z_1(y_{32}+y_{21})y_{32}(w_{12}+w_{23})w_{23}
+x_{32}z_2 y_{21}(y_{32}+y_{21})w_{12}(w_{12}+w_{23})\\
&~~~~~~~~~~~~~~~~~~~~~~+S_{8,\,Y(123)W(321)},\\
&~~~~~~~~~~~~~~~~~~~=(111111)+S_{8,\,Y(123)W(321)},\\
&S_{8,\,Y(123)W(321)}=-\,x_{21}x_{32}z_3(z_2y_{21}w_{12}+z_1y_{32}w_{23})+x_{21}z_3y_{21}y_{32}w_{12}w_{23}
\eal
\ee
for $W(321)$. It is clear that for different orderings of $w$, although their positive variables are different,
the factors corresponding to any contact between boxes are the same. For example, both $W(213)$ and $W(231)$ admit
the third diagram in figure \ref{fig-2}, so the relevant $w$ factors are $w_{12}$ and $(w_{13}\!+\!w_{32})$
respectively, both of which equal to $(w_1\!-\!w_2)$. We also see that $Y(123)W(321)$ admits all six diagrams,
since the orderings of $y$ and $w$ are completely opposite.

Let's sum the six spurious parts over subspaces of $w$ for $Y(123)$, which gives
\be
\bal
S_{Y(123)}=\,&\frac{1}{D_{12}D_{13}D_{23}\,z_1z_2z_3\,w_1w_2w_3}\,X(123)Y(123)
\times x_{21}y_{21}(x_{32}z_2-y_{32}w_2)(z_1w_3-z_3w_1)\\
=\,&(\textrm{prefactors})\times x_{21}y_{21}(x_{32}z_2-y_{32}w_2)(z_1w_3-z_3w_1), \labell{eq-20}
\eal
\ee
and as usual the prefactors are dropped. For the sum of each seed diagram over all subspaces that admit it,
we will present examples of two distinct topologies below.

First, for the first diagram in figure \ref{fig-2}, $x_{21}x_{32}z_1z_2D_{13}$ trivially remains the same
after we sum it over subspaces of $y$ and $w$, since the completeness relation gives
\be
\sum_YY(\ldots)\sum_WW(\ldots)=\frac{1}{y_1y_2y_3\,w_1w_2w_3},
\ee
then it becomes $x_2x_3z_1z_2D_{13}$ after we sum it over subspaces of $x$ that admit it, since
\be
\sum_{\textrm{admitting }X}\!\!\!\!\!\!x_{21}x_{32}=X(123)\,x_{21}x_{32}=\frac{1}{x_1}=\frac{1}{x_1x_2x_3}\,x_2x_3,
\ee
and this is the correct answer as one of those in \eqref{eq-8}.

Then, for the third diagram in figure \ref{fig-2}, $x_{31}x_{32}z_1z_2y_{21}w_{12}$ becomes $x_3^2z_1z_2y_2w_1$ since
\be
\sum_{\textrm{admitting }Y}\sum_{\textrm{admitting }W}\!\!\!\!\!\!y_{21}w_{12}=\frac{1}{y_3w_3}\,Y(12)W(21)\,y_{21}w_{12}
=\frac{1}{y_1y_2y_3\,w_1w_2w_3}\,y_2w_1,
\ee
as well as
\be
\sum_{\textrm{admitting }X}\!\!\!\!\!\!x_{31}x_{32}=X(\sg(12)\,3)\,x_{31}x_{32}=\frac{1}{x_1x_2x_3}\,x_3^2,
\ee
where
\be
X(\sg(12)\,3)=X(123)+X(213)=\frac{x_3}{x_1x_2(x_3-x_1)(x_3-x_2)},
\ee
and this is another one in \eqref{eq-8}. The rest four diagrams of different orientations are similar.

\begin{figure}
\begin{center}
\includegraphics[width=0.75\textwidth]{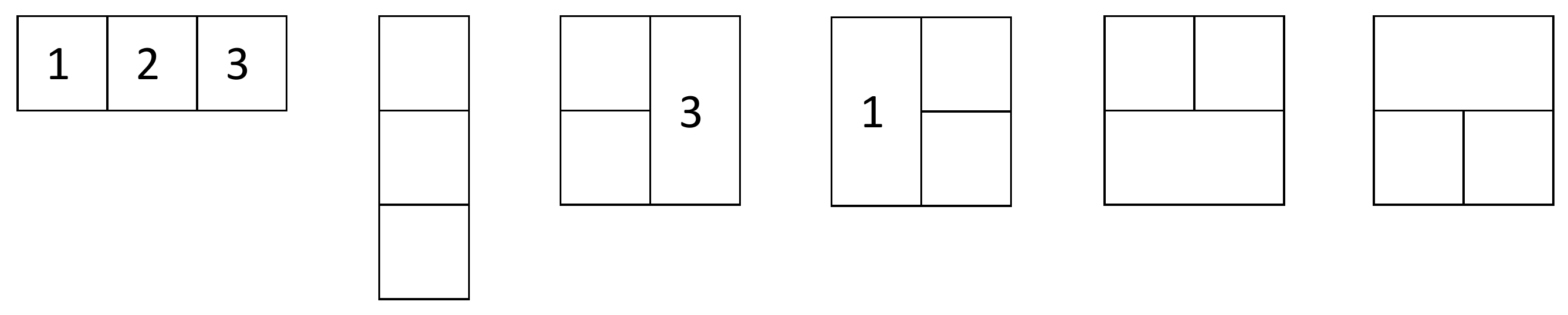}
\caption{Legal seed diagrams in $X(123)$ of which some boxes are kept blank.} \label{fig-3}
\end{center}
\end{figure}

We can continue the separation for the rest five orderings of $y$, each of which contains six orderings of $w$.
Since we still work in $X(123)$, the general seed diagrams for different orderings of $y$ are given in figure \ref{fig-3},
where some boxes are kept blank as the ordering of $x$ alone can only fix part of numbers filled in these boxes.
Straightforwardly, for $Y(132)$ we have
\be
\bal
&T_{2,\,Y(132)W(123)}=(100100)+S_{2,\,Y(132)W(123)},\\
&S_{2,\,Y(132)W(123)}=x_{21}z_1(x_{32}z_2-y_{23}w_{21})y_{31}w_{32}
\eal
\ee
for $W(123)$, and
\be
\bal
&T_{1,\,Y(132)W(132)}=(100000)+S_{1,\,Y(132)W(132)},\\
&S_{1,\,Y(132)W(132)}=0
\eal
\ee
for $W(132)$, and
\be
\bal
&T_{6,\,Y(132)W(213)}=(101110)+S_{6,\,Y(132)W(213)},\\
&S_{6,\,Y(132)W(213)}=x_{21}x_{32}z_1z_2y_{31}w_{31}
\eal
\ee
for $W(213)$, and
\be
\bal
&T_{8,\,Y(132)W(231)}=(111111)+S_{8,\,Y(132)W(231)},\\
&S_{8,\,Y(132)W(231)}=-\,x_{21}z_2(x_{32}z_3-y_{23}w_{32})y_{31}w_{13}
\eal
\ee
for $W(231)$, and
\be
\bal
&T_{3,\,Y(132)W(312)}=(100000)+S_{3,\,Y(132)W(312)},\\
&S_{3,\,Y(132)W(312)}=0
\eal
\ee
for $W(312)$, and
\be
\bal
&T_{7,\,Y(132)W(321)}=(101001)+S_{7,\,Y(132)W(321)},\\
&S_{7,\,Y(132)W(321)}=-\,x_{21}x_{32}z_2z_3y_{31}w_{12}
\eal
\ee
for $W(321)$. The sum of six spurious parts over subspaces of $w$ for $Y(132)$ is
\be
S_{Y(132)}=x_{21}y_{31}(x_{32}z_2(z_1w_3-z_3w_1)-y_{23}w_3(z_1w_2-z_2w_1)). \labell{eq-21}
\ee
\\
Then for $Y(213)$, we have
\be
\bal
&T_{5,\,Y(213)W(123)}=(101000)+S_{5,\,Y(213)W(123)},\\
&S_{5,\,Y(213)W(123)}=x_{21}x_{32}z_1z_2y_{31}w_{21}
\eal
\ee
for $W(123)$, and
\be
\bal
&T_{6,\,Y(213)W(132)}=(101101)+S_{6,\,Y(213)W(132)},\\
&S_{6,\,Y(213)W(132)}=x_{21}x_{32}z_1z_2y_{31}w_{31}
\eal
\ee
for $W(132)$, and
\be
\bal
&T_{1,\,Y(213)W(213)}=(100000)+S_{1,\,Y(213)W(213)},\\
&S_{1,\,Y(213)W(213)}=0
\eal
\ee
for $W(213)$, and
\be
\bal
&T_{3,\,Y(213)W(231)}=(100000)+S_{3,\,Y(213)W(231)},\\
&S_{3,\,Y(213)W(231)}=0
\eal
\ee
for $W(231)$, and
\be
\bal
&T_{8,\,Y(213)W(312)}=(111111)+S_{8,\,Y(213)W(312)},\\
&S_{8,\,Y(213)W(312)}=-\,x_{21}x_{32}z_1z_3y_{31}w_{13}
\eal
\ee
for $W(312)$, and
\be
\bal
&T_{4,\,Y(213)W(321)}=(100110)+S_{4,\,Y(213)W(321)},\\
&S_{4,\,Y(213)W(321)}=-\,x_{21}x_{32}z_1z_3y_{31}w_{23}
\eal
\ee
for $W(321)$. The sum of six spurious parts over subspaces of $w$ for $Y(213)$ is
\be
S_{Y(213)}=0. \labell{eq-22}
\ee
\\
Then for $Y(231)$, we have
\be
\bal
&T_{7,\,Y(231)W(123)}=(101010)+S_{7,\,Y(231)W(123)},\\
&S_{7,\,Y(231)W(123)}=-\,x_{21}x_{32}z_2z_3y_{13}w_{21}
\eal
\ee
for $W(123)$, and
\be
\bal
&T_{8,\,Y(231)W(132)}=(111111)+S_{8,\,Y(231)W(132)},\\
&S_{8,\,Y(231)W(132)}=-\,x_{21}z_2(x_{32}z_3-y_{32}w_{23})y_{13}w_{31}
\eal
\ee
for $W(132)$, and
\be
\bal
&T_{3,\,Y(231)W(213)}=(100000)+S_{3,\,Y(231)W(213)},\\
&S_{3,\,Y(231)W(213)}=0
\eal
\ee
for $W(213)$, and
\be
\bal
&T_{1,\,Y(231)W(231)}=(100000)+S_{1,\,Y(231)W(231)},\\
&S_{1,\,Y(231)W(231)}=0
\eal
\ee
for $W(231)$, and
\be
\bal
&T_{6,\,Y(231)W(312)}=(101101)+S_{6,\,Y(231)W(312)},\\
&S_{6,\,Y(231)W(312)}=x_{21}x_{32}z_1z_2y_{13}w_{13}
\eal
\ee
for $W(312)$, and
\be
\bal
&T_{2,\,Y(231)W(321)}=(100100)+S_{2,\,Y(231)W(321)},\\
&S_{2,\,Y(231)W(321)}=x_{21}z_1(x_{32}z_2-y_{32}w_{12})y_{13}w_{23}
\eal
\ee
for $W(321)$. The sum of six spurious parts over subspaces of $w$ for $Y(231)$ is
\be
S_{Y(231)}=-\,x_{21}y_{32}y_{13}w_2(z_1w_1-z_2w_3). \labell{eq-23}
\ee
\\
Then for $Y(312)$, we have
\be
\bal
&T_{4,\,Y(312)W(123)}=(100101)+S_{4,\,Y(312)W(123)},\\
&S_{4,\,Y(312)W(123)}=-\,x_{21}x_{32}z_1z_3y_{13}w_{32}
\eal
\ee
for $W(123)$, and
\be
\bal
&T_{3,\,Y(312)W(132)}=(100000)+S_{3,\,Y(312)W(132)},\\
&S_{3,\,Y(312)W(132)}=0
\eal
\ee
for $W(132)$, and
\be
\bal
&T_{8,\,Y(312)W(213)}=(111111)+S_{8,\,Y(312)W(213)},\\
&S_{8,\,Y(312)W(213)}=-\,x_{21}x_{32}z_1z_3y_{13}w_{31}
\eal
\ee
for $W(213)$, and
\be
\bal
&T_{6,\,Y(312)W(231)}=(101110)+S_{6,\,Y(312)W(231)},\\
&S_{6,\,Y(312)W(231)}=x_{21}x_{32}z_1z_2y_{13}w_{13}
\eal
\ee
for $W(231)$, and
\be
\bal
&T_{1,\,Y(312)W(312)}=(100000)+S_{1,\,Y(312)W(312)},\\
&S_{1,\,Y(312)W(312)}=0
\eal
\ee
for $W(312)$, and
\be
\bal
&T_{5,\,Y(312)W(321)}=(101000)+S_{5,\,Y(312)W(321)},\\
&S_{5,\,Y(312)W(321)}=x_{21}x_{32}z_1z_2y_{13}w_{12}
\eal
\ee
for $W(321)$. The sum of six spurious parts over subspaces of $w$ for $Y(312)$ is
\be
S_{Y(312)}=x_{21}y_{13}x_{32}z_1(z_2w_1-z_3w_3). \labell{eq-24}
\ee
\\
Finally for $Y(321)$, we have
\be
\bal
&T_{8,\,Y(321)W(123)}=(111111)+S_{8,\,Y(321)W(123)},\\
&S_{8,\,Y(321)W(123)}=-\,x_{21}x_{32}z_3(z_2y_{12}w_{21}+z_1y_{23}w_{32})+x_{21}z_3y_{12}y_{23}w_{21}w_{32}
\eal
\ee
for $W(123)$, and
\be
\bal
&T_{7,\,Y(321)W(132)}=(101010)+S_{7,\,Y(321)W(132)},\\
&S_{7,\,Y(321)W(132)}=-\,x_{21}x_{32}z_2z_3y_{12}w_{31}
\eal
\ee
for $W(132)$, and
\be
\bal
&T_{4,\,Y(321)W(213)}=(100101)+S_{4,\,Y(321)W(213)},\\
&S_{4,\,Y(321)W(213)}=-\,x_{21}x_{32}z_1z_3y_{23}w_{31}
\eal
\ee
for $W(213)$, and
\be
\bal
&T_{2,\,Y(321)W(231)}=(100100)+S_{2,\,Y(321)W(231)},\\
&S_{2,\,Y(321)W(231)}=x_{21}z_1(x_{32}z_2-y_{12}w_{32})y_{23}w_{13}
\eal
\ee
for $W(231)$, and
\be
\bal
&T_{5,\,Y(321)W(312)}=(101000)+S_{5,\,Y(321)W(312)},\\
&S_{5,\,Y(321)W(312)}=x_{21}x_{32}z_1z_2y_{12}w_{13}
\eal
\ee
for $W(312)$, and
\be
\bal
&T_{1,\,Y(321)W(321)}=(100000)+S_{1,\,Y(321)W(321)},\\
&S_{1,\,Y(321)W(321)}=x_{21}x_{32}z_1z_2(y_{12}w_{23}+y_{23}w_{12})
\eal
\ee
for $W(321)$. The sum of six spurious parts over subspaces of $w$ for $Y(321)$ is
\be
S_{Y(321)}=x_{21}y_{23}(x_{32}z_1(z_2w_1-z_3w_3)-y_{12}w_3(z_1w_1-z_3w_2)). \labell{eq-25}
\ee
\\
Collecting the six spurious sums for all ordered subspaces of $y$, namely \eqref{eq-20}, \eqref{eq-21}, \eqref{eq-22},
\eqref{eq-23}, \eqref{eq-24} and \eqref{eq-25}, we can further sum them over $y$-space as
\be
S_{123}=x_{21}(-2\,z_1y_2y_3w_2w_3-z_1y_1w_1(y_2w_3+y_3w_2)+z_2y_3w_3(y_1w_2+y_2w_1)+z_3y_2w_2(y_1w_3+y_3w_1)),
\ee
where subscript $123$ denotes this sum belongs to the sector of $X(123)$. In order to obtain the full result, which
is permutation invariant of loop numbers, we calculate the final spurious sum as
\be
S_{123}X(123)+(5\textrm{ permutations of 1,2,3})=0,
\ee
which nicely vanishes as expected. Therefore, the contributing parts indeed form the correct answer \eqref{eq-8}
which includes all six Mondrian diagrams in figure \ref{fig-2} and their permutations. Note the hidden prefactors are
nothing but the reciprocal of `Denominator' in \eqref{eq-26}.

\newpage
\section{Summary: a Mondrian Preamble}

By separating the contributing and the spurious parts of each form in all ordered subspaces
and assigning the former with corresponding Mondrian factors, which follow simple rules given by
\be
\bal
\textrm{horizontal contact: }&(x_j-x_i)(z_i-z_j)\\
\textrm{vertical contact: }&(y_j-y_i)(w_i-w_j)\\
\textrm{no contact: }&D_{ij}=(x_j-x_i)(z_i-z_j)+(y_j-y_i)(w_i-w_j)
\eal
\ee
between any two loops labelled by $i,j$, we obtain the seed diagrams. If we assume the spurious terms will always
sum to zero at the end, there is no need to sum the seed diagrams over all ordered subspaces since they
are already topologically valuable. There is a simple way to find seed diagrams: let's work in simply one
ordered subspace $X(12)Z(21)Y(12)W(21)$ at 2-loop, as the first nontrivial example. Then, it is clear that $D_{12}$
is trivially positive so there is no positivity condition to be imposed. But as a physical pole $D_{12}$ must
appear in the denominator, which identically turns the form into
\be
\frac{1}{x_1x_2z_1z_2y_1y_2w_1w_2}\frac{D_{12}}{D_{12}}
=\frac{1}{x_1x_2z_1z_2y_1y_2w_1w_2}\frac{(x_2-x_1)(z_1-z_2)+(y_2-y_1)(w_1-w_2)}{D_{12}}.
\ee
As usual, dropping the prefactors which contain all physical poles, we precisely obtain two 2-loop ladders of
horizontal and vertical orientations (the vertical one is shown in figure \ref{fig-1}).

The 3-loop example is more interesting. Similarly in ordered subspace $X(123)Z(321)Y(123)W(321)$, we can separate the triple
product as
\be
\bal
D_{12}D_{13}D_{23}=\,&\,x_{21}z_{12}\cdot x_{32}z_{23}\cdot D_{13}+y_{21}w_{12}\cdot y_{32}w_{23}\cdot D_{13}\\
&+x_{31}z_{13}\cdot x_{32}z_{23}\cdot y_{21}w_{12}+x_{21}z_{12}\cdot x_{31}z_{13}\cdot y_{32}w_{23}\\
&+x_{21}z_{12}\cdot y_{31}w_{13}\cdot y_{32}w_{23}+y_{21}w_{12}\cdot y_{31}w_{13}\cdot x_{32}z_{23},
\eal
\ee
which precisely correspond to the six diagrams in figure \ref{fig-2} (including two ladders and four tennis courts).
Here, for notational compactness we have defined $x_{31}\!\equiv\!x_{32}\!+\!x_{21}$ for instance,
as $x_{32}$ and $x_{21}$ are primitive positive variables in this subspace while $x_{31}$ is not.

In general, Mondrian diagrams of higher loop orders satisfy this neat pattern: the product of all $D_{ij}$'s
can be expanded as a sum of all topologies and orientations in an ordered subspace of which the orderings of $x,z$
are completely opposite, so are those of $y,w$. However, there are more subtle issues to be clarified,
and we will continue to discuss them more systematically in the subsequent work.

\section*{Acknowledgments}

This work is partly supported by Qiu-Shi Funding and Chinese NSF funding under contracts\\
No.11135006, No.11125523 and No.11575156.

\newpage
\appendix
\section{The Master Form at 4-loop}
\label{app1}

To obtain the master form $T_{64}$ (its subscript is due to $2^6\!=\!64$) at 4-loop by dividing the $z$-space, we define
\be
\bal
\eta_{12}\equiv z_1-z_2+c_{12}>0,~~\eta_{13}\equiv z_1-z_3+c_{13}>0,~~\eta_{23}\equiv z_2-z_3+c_{23}>0,\\
\eta_{14}\equiv z_1-z_4+c_{14}>0,~~\eta_{24}\equiv z_2-z_4+c_{24}>0,~~\eta_{34}\equiv z_3-z_4+c_{34}>0,
\eal
\ee
the sum is then
\be
\bal
T_{64}\equiv\,&~\,Z^-_{14}\cap Z^-_{24}\cap Z^-_{34}\cap Z^-_{13}\cap Z^-_{23}\cap Z^-_{12}\\
=\,&~\,\frac{1}{c_{12}}\[~~~\frac{1}{c_{13}c_{23}}\(\frac{1}{c_{14}c_{24}c_{34}}\,Z(4321)
+\frac{1}{c_{14}c_{24}\eta_{34}}\,Z(3421)+\frac{1}{c_{14}\eta_{24}\eta_{34}}\,Z(3241)
+\frac{1}{\eta_{14}\eta_{24}\eta_{34}}\,Z(3214)\)\right.\\
&~~~~~~~\,+\frac{1}{c_{13}\eta_{23}}\(\frac{1}{c_{14}c_{24}c_{34}}\,Z(4231)
+\frac{1}{c_{14}\eta_{24}c_{34}}\,Z(2431)+\frac{1}{c_{14}\eta_{24}\eta_{34}}\,Z(2341)
+\frac{1}{\eta_{14}\eta_{24}\eta_{34}}\,Z(2314)\)\\
&~~~~~~~\,+\!\left.\frac{1}{\eta_{13}\eta_{23}}\(\frac{1}{c_{14}c_{24}c_{34}}\,Z(4213)
+\frac{1}{c_{14}\eta_{24}c_{34}}\,Z(2413)+\frac{1}{\eta_{14}\eta_{24}c_{34}}\,Z(2143)
+\frac{1}{\eta_{14}\eta_{24}\eta_{34}}\,Z(2134)\)\]\\
&\!\!\!\!+\frac{1}{\eta_{12}}\[~~~\frac{1}{c_{13}c_{23}}\(\frac{1}{c_{14}c_{24}c_{34}}\,Z(4312)
+\frac{1}{c_{14}c_{24}\eta_{34}}\,Z(3412)+\frac{1}{\eta_{14}c_{24}\eta_{34}}\,Z(3142)
+\frac{1}{\eta_{14}\eta_{24}\eta_{34}}\,Z(3124)\)\right.\\
&~~~~~~~\,+\frac{1}{\eta_{13}c_{23}}\(\frac{1}{c_{14}c_{24}c_{34}}\,Z(4132)
+\frac{1}{\eta_{14}c_{24}c_{34}}\,Z(1432)+\frac{1}{\eta_{14}c_{24}\eta_{34}}\,Z(1342)
+\frac{1}{\eta_{14}\eta_{24}\eta_{34}}\,Z(1324)\)\\
&~~~~~~~\,+\!\left.\frac{1}{\eta_{13}\eta_{23}}\(\frac{1}{c_{14}c_{24}c_{34}}\,Z(4123)
+\frac{1}{\eta_{14}c_{24}c_{34}}\,Z(1423)+\frac{1}{\eta_{14}\eta_{24}c_{34}}\,Z(1243)
+\frac{1}{\eta_{14}\eta_{24}\eta_{34}}\,Z(1234)\)\]\\
=\,&~\,\frac{1}{c_{12}c_{13}c_{14}c_{23}c_{24}c_{34}}
\frac{N}{z_1z_2z_3z_4\,\eta_{12}\eta_{13}\eta_{14}\eta_{23}\eta_{24}\eta_{34}},
\eal
\ee
where
\be
\bal
N=\,\,&(z_1+c_{12})(z_1+c_{13})(z_1+c_{14})(z_2+c_{23})(z_2+c_{24})(z_3+c_{34})\\
&-z_1z_2z_3(z_1+c_{14})(z_2+c_{24})(z_3+c_{34})-z_1z_2z_4(z_1+c_{13})(z_2+c_{23})(z_3+c_{34})\\
&-z_1z_3z_4(z_1+c_{12})(z_2+c_{23})(z_2+c_{24})-z_2z_3z_4(z_1+c_{12})(z_1+c_{13})(z_1+c_{14})\\
&+z_1z_2z_3z_4\(-\,(z_1+c_{12})(z_3+c_{34})-(z_1+c_{13})(z_2+c_{24})-(z_1+c_{14})(z_2+c_{23})\right.\\
&~~~~~~~~~~~~~~~\,+(z_1+c_{12})(z_3+z_4)+(z_1+c_{13})(z_2+z_4)+(z_1+c_{14})(z_2+z_3)\\
&~~~~~~~~~~~~~~~\,+(z_2+c_{23})(z_1+z_4)+(z_2+c_{24})(z_1+z_3)+(z_3+c_{34})(z_1+z_2)\\
&~~~~~~~~~~~~~~~\,\left.-\,z_1z_2-z_1z_3-z_1z_4-z_2z_3-z_2z_4-z_3z_4\). \labell{eq-27}
\eal
\ee
Now, let's determine $T_1\!\equiv\!Z^+_{41}\cap Z^+_{42}\cap Z^+_{43}\cap Z^+_{31}\cap Z^+_{32}\cap Z^+_{21}$,
for instance, by flipping all $c_{ij}$'s to $-c_{ji}$'s in the denominator and
setting all $c_{ij}$'s to zero in the numerator, which gives
\be
T_1=\frac{1}{c_{12}c_{13}c_{14}c_{23}c_{24}c_{34}}
\frac{(z_1-z_3)(z_1-z_4)(z_2-z_4)}{z_4\,\zeta_{12}\zeta_{13}\zeta_{14}\zeta_{23}\zeta_{24}\zeta_{34}}, \labell{eq-29}
\ee
where we have similarly defined
\be
\bal
\zeta_{12}\equiv z_1-z_2-c_{21}>0,~~\zeta_{13}\equiv z_1-z_3-c_{31}>0,~~\zeta_{23}\equiv z_2-z_3-c_{32}>0,\\
\zeta_{14}\equiv z_1-z_4-c_{41}>0,~~\zeta_{24}\equiv z_2-z_4-c_{42}>0,~~\zeta_{34}\equiv z_3-z_4-c_{43}>0.
\eal
\ee
To confirm this is indeed the correct answer, we can separate it into two parts as
\be
T_1=\(Z^+_{41}\cap Z^+_{42}\cap Z^+_{31}\)\cap\(Z^+_{43}\cap Z^+_{32}\cap Z^+_{21}\)
=\frac{(z_1-z_4)(z_2-z_4)(z_1-z_3)}{c_{14}c_{24}c_{13}\,\zeta_{14}\zeta_{24}\zeta_{13}}\times
\frac{1}{c_{34}c_{23}c_{12}\,z_4\,\zeta_{34}\zeta_{23}\zeta_{12}},
\ee
where the second part can be trivially obtained if we treat $\zeta_{34},\zeta_{23},\zeta_{12}$ as genuinely
positive variables. For the first part, in terms of $\zeta_{34},\zeta_{23},\zeta_{12}$ we have
\be
\zeta_{13}=\zeta_{12}+\zeta_{23}+c_{21}+c_{32}-c_{31}>0\Longrightarrow c_{31}<\zeta_{12}+\zeta_{23}+c_{21}+c_{32},
\ee
and the corresponding form is
\be
\frac{1}{c_{31}}-\frac{1}{c_{31}-(\zeta_{12}+\zeta_{23}+c_{21}+c_{32})}=\frac{z_1-z_3}{c_{13}\,\zeta_{13}}.
\ee
Analogously we have
\be
\zeta_{24}=\zeta_{23}+\zeta_{34}+c_{32}+c_{43}-c_{42}>0\Longrightarrow\frac{z_2-z_4}{c_{24}\,\zeta_{24}},
\ee
as well as
\be
\zeta_{14}=\zeta_{12}+\zeta_{23}+\zeta_{34}+c_{21}+c_{32}+c_{43}-c_{41}>0\Longrightarrow\frac{z_1-z_4}{c_{14}\,\zeta_{14}},
\ee
therefore, we have neatly confirmed $T_1$'s expression \eqref{eq-29}.

Another check of the master form is the completeness relation
\be
Z^-_{14}\cap Z^-_{24}\cap Z^-_{34}\cap Z^-_{13}\cap Z^-_{23}\cap Z^-_{12}
+Z^-_{14}\cap Z^-_{24}\cap Z^+_{34}\cap Z^-_{13}\cap Z^-_{23}\cap Z^-_{12}
=Z^-_{14}\cap Z^-_{24}\cap Z^-_{13}\cap Z^-_{23}\cap Z^-_{12},~
\ee
for instance, of which the essential part is $Z^-_{34}\!+\!Z^+_{34}\!=\!I_{34}$. To prove this,
we can focus on the numerators while omitting their common denominator, so that the relation becomes
\be
N-N(c_{34}\!=\!0,3\!\leftrightarrow\!4)=\(\frac{N}{\eta_{34}}\)_{c_{34}\to\infty}\eta_{34}, \labell{eq-30}
\ee
where $N$ is the numerator in \eqref{eq-27}. Let's immediately give some explanation of the quantities above.
For the RHS, unlike the 3-loop case \eqref{eq-28}, it is much more nontrivial to fix this quintuple co-positive product.
Hence we use another way to circumvent it, which is extremely simple: we set $c_{34}\!\to\!\infty$ and then evaluate its
residue, since this trivializes $z_3\!-\!z_4\!+\!c_{34}\!>\!0$, after that we need a compensating factor $\eta_{34}$
as terms of the LHS are purely numerators. For the second term of the LHS, the operation
$(c_{34}\!=\!0,3\!\leftrightarrow\!4)$ is easy to understand, while the minus sign comes from
the operation $(c_{34}\!\to\!-c_{43},3\!\leftrightarrow\!4)$ of the denominator
\be
\frac{1}{z_3+c_{34}-z_4}\to\frac{1}{z_3-c_{43}-z_4}\to\frac{1}{z_4-c_{34}-z_3}=-\frac{1}{z_3+c_{34}-z_4},
\ee
as this term demands $z_4\!-\!z_3\!-\!c_{34}\!>\!0$, but to have a common denominator produces an minus sign.
And a simple explicit calculation confirms \eqref{eq-30}, which finishes the check.


\end{document}